\documentclass[reprint, superscriptaddress, amsmath, amssymb, aps, prd, nofootinbib,longbibliography]{revtex4-1}
\usepackage{graphicx}
\usepackage[caption=false]{subfig}
\usepackage{bm}
\usepackage{hyperref}
\hypersetup{
    colorlinks=true,
    linkcolor=blue,
    filecolor=magenta,      
    urlcolor=blue,
    citecolor=blue
    }
\usepackage{bm}
\usepackage{braket}
\usepackage{listings}
\usepackage{etoolbox}
\usepackage{amsfonts}
\usepackage{soul}
\usepackage{commath}
\usepackage[capitalise,nameinlink]{cleveref} 
\usepackage{siunitx}
\usepackage{color}
\newcommand{\be}{ \begin{equation}}
\newcommand{\ee}{\end{equation}}
\renewcommand{\vec}[1]{\mathbf{#1}}
\DeclareSIUnit\parsec{pc}
\DeclareSIUnit\planckmass{m_\mathrm{p}}

\begin{document}
\title{Quantum initial conditions for inflation and canonical invariance}

\author{F.J. Agocs}
\email{fa325@cam.ac.uk}
\affiliation{Astrophysics Group, Cavendish Laboratory, J.J.Thomson Avenue, Cambridge, CB3 0HE, UK}
\affiliation{Kavli Institute for Cosmology, Madingley Road, Cambridge, CB3 0HA, UK}
\author{L.T. Hergt}
\email{lh561@cam.ac.uk}
\affiliation{Astrophysics Group, Cavendish Laboratory, J.J.Thomson Avenue, Cambridge, CB3 0HE, UK}
\affiliation{Kavli Institute for Cosmology, Madingley Road, Cambridge, CB3 0HA, UK}
\author{W.J. Handley}
\email{wh260@cam.ac.uk}
\affiliation{Astrophysics Group, Cavendish Laboratory, J.J.Thomson Avenue, Cambridge, CB3 0HE, UK}
\affiliation{Kavli Institute for Cosmology, Madingley Road, Cambridge, CB3 0HA, UK}
\author{A.N. Lasenby}
\email{a.n.lasenby@mrao.cam.ac.uk}
\affiliation{Astrophysics Group, Cavendish Laboratory, J.J.Thomson Avenue, Cambridge, CB3 0HE, UK}
\affiliation{Kavli Institute for Cosmology, Madingley Road, Cambridge, CB3 0HA, UK}
\author{M.P. Hobson}
\email{mph@mrao.cam.ac.uk}
\affiliation{Astrophysics Group, Cavendish Laboratory, J.J.Thomson Avenue, Cambridge, CB3 0HE, UK}
\date{\today}

\begin{abstract}

We investigate the transformation of initial conditions for primordial curvature perturbations under two types of transformations of the associated action: simultaneous redefinition of time and the field to be quantised, and the addition of surface terms. The latter encompasses all canonical transformations, whilst the time- and field-redefinition is a distinct, non-canonical transformation since the initial and destination systems use different times.
Actions related to each other via such transformations yield identical equations of motion and preserve the commutator structure. They further preserve the time-evolution of expectation values of quantum operators unless the vacuum state also changes under the transformation. These properties suggest that it is of interest to investigate vacuum prescriptions that also remain unchanged under canonical transformations.
We find that initial conditions derived via minimising the vacuum expectation value of the Hamiltonian and those obtained using the Danielsson vacuum prescription are \emph{not} invariant under these transformations, whereas those obtained by minimising the local energy density are invariant. We derive the range of physically distinct initial conditions obtainable by Hamiltonian diagonalisation, and illustrate their effect on the scalar primordial power spectrum and the Cosmic Microwave Background under the `just enough inflation' model. We also generalise the analogy between the dynamics of a quantum scalar field on a curved, time-dependent spacetime and the gauge-invariant curvature perturbation.
We argue that the invariance of the vacuum prescription obtained by minimising the renormalised stress--energy tensor should make it the preferred procedure for setting initial conditions for primordial perturbations. All other procedures reviewed in this work yield ambiguous initial conditions, which is problematic both in theory and in practice.
\end{abstract}

\maketitle
\section{Introduction}\label{sec:intro}

Anisotropies in the Cosmic Microwave Background (CMB) are thought to have been seeded during cosmic inflation by quantum fluctuations.  To model the statistical properties of the CMB, one therefore needs either to assume a spectrum of such perturbations at the end of inflation, or compute one. To compute a spectrum, the relevant equations one needs to solve are the cosmological field equations \cite{friedmann}, which describe the time-evolution of a homogeneous, expanding `background' universe, alongside the Mukhanov--Sasaki equation \cite{mukhanov}, a second-order ordinary differential equation which governs the growth of small fluctuations in the metric and matter field(s) on this background. The gauge-invariant curvature perturbations `freeze out' during inflation. They are therefore evolved numerically until well into inflation, and their frozen-out amplitudes are used to compute the primordial power spectrum. The numerical solution requires two initial conditions for the Fourier modes of perturbations (referred to as mode functions), which are usually motivated by quantum mechanical vacuum considerations. Often the initial quantum state for the primordial perturbations is chosen so as to minimise the Hamiltonian density. In an expanding spacetime, however, the Hamiltonian becomes time-dependent, leading to the ground state at a given time no longer being the ground state at a later time. The divergent Hamiltonian then yields infinite particle density at times other than the instant the initial state is set \cite{Fulling1979, nqicfi}. \citet{nqicfi} thus suggest minimisation of a quantity describing the local energy density instead, and derive a different set of vacuum conditions. The above both define the initial state as some minimal energy density state, but there are many others that instead define it as the state annihilated by annihilation operators, such as one proposed by \citet{bunch-davies-vacuum}, by \citet{danielsson}, the $\alpha$-vacua \cite{chernikov-tagirov,tagirov,allen,mottola, goldstein-lowe-alphavacua}, and others. 

In this work we review the aforementioned procedures for setting initial conditions from the perspective of invariance under a set of transformations that are canonical.
Canonical transformations leave the classical evolution and the commutator structure of the system invariant by construction.
The behaviour of expectation values of quantum operators under canonical transformations is not as obvious: \citet{matacz,bozza} both argued that since the wavefunction only changes by a field-dependent phase, the expectation values must be invariant, but while \citet{matacz} concludes that vacua arising from different choices of canonical variables are equivalent, \citet{bozza} observes that the vacuum (the state that minimises the vacuum expectation value of the Hamiltonian) does in fact change. \citet{venningrain2019} then proved formally that if the two sets of canonical variables select out different vacuum states (as is the case with Hamiltonian diagonalisation), then there will be differences in expectation values and the observable differences depend only on the canonical transformation at the time when the initial state (e.g. the vacuum) is set. If canonical variables yield the same initial (vacuum) state, then the expectation values of operators will also match.
This suggests that it is of interest to investigate vacuum prescriptions that also remain unchanged under canonical transformations. Such a prescription would unambiguously define initial conditions for the perturbation mode functions, and thus would be the preferred choice for setting the vacuum. This is because any ambiguity arising from other choices in the initial conditions, which do not share this invariance, would lead to uncertainty in the frozen-out amplitude of the mode functions, the primordial power spectrum, and ultimately in the power spectrum of CMB anisotropies.

As first observed by \citet{Fulling1979} in 1979, the initial conditions resulting from the popular Hamiltonian diagonalisation method (minimising the Hamiltonian density) do in fact change under canonical transformations, a fact that he used to argue against this procedure and the reason that we consider instead minimising the local energy density via the renormalised stress--energy tensor preferable over the Hamiltonian diagonalisation approach to setting initial conditions.
Weiss \cite{weiss} then added field-redefinitions to the group of canonical transformations considered by Fulling, but instead came to the conclusion that a preferred Hamiltonian can be picked out for its desirable mathematical properties.  This preferred Hamiltonian is the one written in terms of conformal time $\eta$ and the Mukhanov variable $v$. The preferred Hamiltonian coincides with the one considered conventionally, for its action describes a canonically normalised scalar field. 
A recent study \cite{venningrain2019} provides a thorough review of canonical transformations and their effect on scalar field fluctuations during inflation, observing that the transformations can be used to select out different vacuum states.

In our current work, in addition to reinforcing Fulling's findings that the Hamiltonian diagonalisation vacuum is a canonically non-invariant procedure and demonstrating how this property may affect observations, we show that the Danielsson vacuum suffers the same pathology, whereas minimising the local energy density through the renormalised stress--energy tensor does not. 
The latter two procedures were not available to Fulling at the time. In spite of these potential theoretical shortcomings of the Hamiltonian diagonalisation and Danielsson vacua, amongst others, it should be pointed out that, in practice, these methods for setting the vacuum state for primordial perturbations will lead to the same observable consequences as minimising the local energy density if they are applied at an epoch
when the modes in question lie well within the Hubble horizon. Nonetheless, as we will show, observable differences do occur in particular for `just enough inflation' models. Such models may be particularly relevant for closed cosmologies  \cite{Bonga2016,Uzan2003,Ellis2002,Ellis2002-causality,Lasenby-closed}, but we restrict our attention in this work to spatially flat spacetimes for simplicity.

\cref{sec:bg} covers the relevant mathematical and physical background, with \cref{sec:bg-cosmo} summarising the classical theory of inflationary perturbations and \cref{sec:bg-vacua} reviewing how the vacuum choices considered arise. 
\cref{sec:methods} then explains what transformations of the vacuum-setting procedures are carried out in this work. Results are presented in \cref{sec:results}, broken into subsections to show the effect of the canonical transformations considered, under each vacuum prescription, on the initial conditions, the primordial power spectrum of scalar curvature perturbations, and the temperature-temperature power spectrum of fluctuations in the CMB, respectively, under the `just enough inflation' model. We draw conclusions in \cref{sec:conclusion}.

The notation used throughout this paper is summarised in \cref{table:notation}. We use reduced Planck units such that $\hbar=c=k_{\text{B}}=8\pi G=1$. Greek indices can have any value from $0$ to $3$, with $0$ reserved for time and $1$--$3$ denoting spatial components. Latin indices indicate spatial components. We adopt the metric signature $(+,-,-,-)$.

\begin{table}[htb]
\renewcommand{\arraystretch}{1.5}
\begin{ruledtabular}
\caption{\label{table:notation} Summary of our notation.}
\begin{tabular}{cc}
symbol  & meaning      \\\hline
$t$ & cosmic time \\
$\eta$ & conformal time \\ 
$\tau$ & an arbitrary timelike independent variable \\
$\dot{f}$ & $\frac{\partial f}{\partial t}$ \\
$f'$ & $\frac{\partial f}{\partial \eta}$ \\
$\partial_{\tau} f$ & $\frac{\partial f}{\partial \tau}$ 
\end{tabular}
\end{ruledtabular}
\end{table}

\section{Background}\label{sec:bg}
 
\subsection{Canonical transformations}\label{sec:bg-canonical}

To motivate the choice of transformations considered in this work, we briefly review `classical' canonical transformations and a broader class of canonical transformations performed on an extended phase space. For a thorough review of canonical transformations in cosmology, see \cite{venningrain2019}.

In the Lagrangian formulation of classical mechanics, all information about the dynamics of the system is carried by the Lagrangian $L(q,\dot{q},t)$, a function of the $n$ generalised coordinates $q_i$ and their derivatives. The Euler--Lagrange equations yield the equations of motion even if one performs a reversible change of coordinates (or point transformation)
\begin{equation}\label{eq:lagrangian-coord-change}
q_i \rightarrow Q_i(q,t).
\end{equation}

The Hamiltonian formulation of mechanics puts the generalised coordinates $q_i$ and their derivatives $\dot{q}_i$ on an equal footing by considering generalised momenta 
\begin{equation}
p_i  = \frac{\partial L}{\partial \dot{q}_i}
\end{equation}
and working in the $2n$-dimensional phase space ($q_i$,$p_i$) instead of configuration space ($q_i,\dot{q}_i$). In the Hamiltonian formalism there exists a broader range of transformations than (\ref{eq:lagrangian-coord-change}) that leave the Hamiltonian equations of motion invariant, called \emph{canonical transformations}. Canonical transformations allow one to mix the generalised coordinates and momenta and to do so in a time-dependent manner
\be 
q_i \rightarrow Q_i(q_i, p_i, t), \quad p_i \rightarrow P_i(q_i, p_i, t),  
\ee
whilst preserving the form of Hamilton's equations of motion,
\be
\begin{gathered}
\dot{p}_i = -\frac{\partial H}{\partial q_i}, \quad \dot{q}_i = \frac{\partial H}{\partial p_i}, \\
\dot{P}_i = -\frac{\partial K}{\partial Q_i}, \quad \dot{Q}_i = \frac{\partial K}{\partial P_i},
\end{gathered}
\ee
where $K$ is the new Hamiltonian. For the equations of motion to be conserved, the principle of extremal action has to be satisfied in both the initial and transformed system:
\begin{align}
\delta \int_{t_1}^{t_2} (p_i\dot{q}_i - H(q,p,t))dt &= 0,\label{eq:leastaction-ini} \\
\delta \int_{t_1}^{t_2} (P_i\dot{Q}_i - K(Q,P,t))dt &= 0. \label{eq:leastaction-final}
\end{align}
For (\ref{eq:leastaction-ini}) to imply (\ref{eq:leastaction-final}), we require
\be\label{eq:total-diff-lagrangian}
\lambda(p_i\dot{q}_i - H) = P_i\dot{Q}_i - K + \dot{F},
\ee
with $\lambda = 1$\footnote{Transformations with $\lambda\neq1$ are called extended canonical transformations, and will not be considered in this work.}. The Lagrangian of the system can at most change by a total derivative $dF$, with $F$ being called the generating function of the transformation. All canonical transformations possess a generating function \cite{lanczos-mechanics,calkin-mechanics}.
The Hamiltonian then transforms as
\be\label{eq:Hamiltonian-change}
K = H + \frac{\partial F}{\partial t}.
\ee
It can be shown that the Poisson bracket, 
\begin{align}
\{f,g\} = \sum_{i=1}^{N} \frac{\partial f}{\partial q_i}\frac{\partial
g}{\partial p_i} - \frac{\partial f}{\partial p_i}\frac{\partial g}{\partial q_i}, 
\end{align}
is invariant under canonical transformations, and that the converse is true: if the Poisson bracket structure is conserved such that 
\be\label{eq:poisson-br} 
\{Q_i, Q_j\} = \{P_i, P_j\} = 0, \quad \{Q_i, P_j\} = \delta_{ij}, 
\ee
then the transformation is canonical. 

The (quantum) field theoretical equivalent of (\ref{eq:total-diff-lagrangian}), adding a total differential to the Lagrangian, is the freedom to add a 4-divergence to the Lagrangian density, 
\be 
\mathcal{L}' = \mathcal{L}(t,\phi, \partial_{\mu}\phi) + \nabla_{\mu}J^{\mu}(\phi, \partial_{\mu}\phi),
\ee
which, if the independent variable stays the same throughout, is equivalent to integrating the associated action by parts and discarding the surface term. 
One of the transformations considered in this work is the addition of such vanishing surface terms, which thus covers all canonical transformations.
Commutators are the quantum counterparts of Poisson brackets, and are thus preserved under canonical transformations. In all transformations described in this work, it is ensured that the commutator structure of the destination system matches that of the original.

In the conventional formulation of the principle of least action, $t$ is the Newtonian absolute time, and has a distinguished role. This is not always desirable and for relativistic considerations, one may wish to treat the generalised coordinates $q_i$ and $t$ on an equal footing. This can be achieved by extending the phase space considered with time $t$ and its conjugate momentum, the negative Hamiltonian: $(q_i, p_i, t, -\mathcal{H})$ \cite{Tsyganov,Greiner,struckmeier-riedel}. In this extended space it is possible to define an extended set of Hamilton's equations, and define canonical transformations that involve a redefinition of time,
\be
(t, q_i, p_i, -\mathcal{H}) \rightarrow (T, Q_i, P_i, -\mathcal{K}).
\ee
The other type of transformations we consider in this work, namely simultaneous time- and field-redefinitions, are canonical transformations on the extended phase space. While in itself it does not include all canonical transformations on the extended space, it is motivated by practical considerations: one may wish to derive initial conditions `from first principles' for a different field, related to the original one by a time-dependent rescaling, for its better stability properties in numerical simulations.

\subsection{Dynamics of primordial perturbations}\label{sec:bg-cosmo}

\subsubsection{The perturbed classical action}\label{sec:semiclassical-pert}

The dynamics of primordial curvature perturbations arise from perturbing the metric and matter fields around a homogeneous, isotropic, expanding background (for a thorough review, see e.g.\ \cite{mukhanov, Maldacena, tasi}).  In an inflationary model with a single scalar field $\phi(t)$ and potential $V(\phi)$ on a Friedmann--Robertson--Walker (FRW) spacetime, the system can be described via the action
\be\label{eq:classical-action}
S = \tfrac{1}{2}\int \mathop{d^4x} \sqrt{-g}
\left[R - (\partial_{\mu}\phi)(\partial^{\mu}\phi) - 2V(\phi)\right],
\ee
where the metric $g_{\mu \nu}$ is the spatially flat FRW metric.  Perturbing this action in the comoving gauge 
\be\label{eq:gauge-choice} 
\delta \phi = 0, \: g_{ij} = a^2[(1-2\mathcal{R})\delta_{ij} + h_{ij}], \: \partial_i h_{ij} =h_i^i = 0, 
\ee
yields, to second order in the gauge-invariant curvature perturbation $\mathcal{R}$, 
\be\label{eq:pert-action}
S_2 = \tfrac{1}{2}\int \mathop{d^4x} az^2\left[ \dot{\mathcal{R}}^2 - a^{-2}(\partial_i\mathcal{R})^2 \right].
\ee
Traditionally this is then written in terms of the Mukhanov variable $v = z\mathcal{R} = \frac{a\dot{\phi}}{H}\mathcal{R}$ and conformal time $\eta$ and integrated by parts to give 
\be\label{eq:action-mukhanov} 
S_2 = \tfrac{1}{2}\int \mathop{d\eta}\mathop{d^3x} \Big[ (v')^2 - (\partial_i v)^2 + \frac{z''}{z}v^2 \Big]. 
\ee
The reason behind the variable choice and integration by parts is that in the resulting action $v$ is canonically normalised, i.e.\ it has a kinetic term of the form $\frac{1}{2}\partial_{\mu}v \partial^{\mu}v$, and thus yields an equation of motion of a particular form. Writing $v$ as a Fourier decomposition,
\be\label{eq:fourier-v}
v(\eta, x) = \int \frac{\mathop{d^3 k}}{(2\pi)^3} v_{\vec{k}}(\eta)
e^{i\vec{k}\cdot\vec{x}},
\ee
the action (\ref{eq:action-mukhanov}) gives the equation of motion of an oscillator: 
\be\label{eq:ms-v}
v_k'' + \left(k^2 - \frac{z''}{z} \right)v_k = 0,
\ee
where the vector notation on the wavevector $k$ has been removed due to the isotropy of the field $v$. The variables $(\eta, v)$ were thus chosen because they yield an oscillator's equation of motion (without a `first-derivative term' proportional to $v'$), allowing the classical field $v$ to be quantised by analogy with a time-dependent quantum harmonic oscillator.

\subsubsection{Scalar fields in curved spacetime}\label{sec:scalar-fields}
 
In order to obtain the action (\ref{eq:pert-action}) describing the dynamics of the gauge-invariant curvature perturbation, action (\ref{eq:classical-action}) was perturbed in both the metric and the inflaton field, but a gauge was then chosen to keep only the metric perturbations and leave the inflaton unperturbed. One can arrive at analogous dynamics (a similar equation of motion to (\ref{eq:ms-v})) by instead modelling the field perturbations as a new massless scalar field $\varphi$ on a spacetime with an unperturbed metric.
Starting from the action describing the dynamics of the field perturbations $\varphi$,
\be\label{eq:action-scalar}
S = \tfrac{1}{2} \int \mathop{d^4x} \sqrt{-g} \left( g^{\mu
\nu} \partial_{\mu} \varphi \partial_{\nu} \varphi - m^2\varphi^2 \right),
\ee
which one traditionally considers in terms of the auxiliary field $y = a\varphi$ and conformal time (for similar reasons as $v$ was considered in the previous section), one derives the Fourier space equation of motion 
\be\label{eq:eom-scalar}
y_k'' + \left( k^2 - \frac{a''}{a} + a^2m^2 \right)y_k = 0.
\ee
The similarity between (\ref{eq:ms-v}) and (\ref{eq:eom-scalar}) means that one can \emph{identically} map the modes $y_k$ derived from a massless scalar field $\varphi$ onto the perturbation modes $v_k$, provided one ensures that the two associated background spacetimes satisfy
\be\label{eq:equivalence-cond}
\frac{a''}{a} = \frac{z''}{z}. 
\ee
One particular solution of (\ref{eq:equivalence-cond}) is the case $z \propto a$, which holds true or becomes a good approximation in a number of cases. It is true exactly in power-law inflation models \cite{pli-1}, and becomes a good approximation in slow-roll inflation \cite{martin-brandenberger-transplanckian-1}, which is power-law inflation to first order and is admitted by many models. The proportionality is also a good approximation in kinetic dominance \cite{kineticic}, which we describe in more detail in \cref{sec:results-pps}. Therefore, in these cases, one may treat the Mukhanov variable in an inflating universe as if it were a scalar field $\varphi(\eta, x) /a(\eta)$ on the same background, obeying the action
(\ref{eq:action-scalar}). It should be emphasised that one may always map the results for the test scalar field onto the primordial perturbations, but in general the mapping will not be the identity and so the equations of motion will not be form-identical. The advantage of working with action (\ref{eq:action-scalar}) over (\ref{eq:action-mukhanov}) is that the former is in covariant form (once each $\partial_{\mu}$ has been replaced by $\nabla_{\mu}$), and hence can be used to derive further covariant quantities such as the stress--energy tensor, the minimisation of which provides a definition of the ground state (see \cref{sec:classical-rst}). 

\subsection{Vacuum choices}\label{sec:bg-vacua}

In the above models of primordial perturbations there are multiple ways to define a vacuum or ground state. These result in expressions for the Fourier modes of the perturbations considered ($v_k$ or $y_k$), which can then be used as initial conditions for the perturbation modes, and affect the form of the primordial power spectrum the modes admit. In this section we introduce three different definitions of the vacuum state: \emph{Hamiltonian diagonalisation}, the \emph{Danielsson vacuum}, and \emph{minimising the renormalised stress--energy tensor}. We derive the initial conditions each vacuum definition gives for the mode functions if applied in the conventional way. We then examine how the initial conditions change under a field-redefinition in the associated action and the addition of surface terms in \cref{sec:results-ic}. 

Acknowledging that the literature on inflationary initial conditions is vast, in \cref{sec:other-vacua} we additionally list the Bunch-Davies vacuum, the adiabatic vacuum and $\alpha$-vacua. The behaviour of some of these vacua under canonical transformations has been investigated in the past, but for others this is either impossible (due to them relying on a special choice of canonical variable pair) or remains to be carried out.

\subsubsection{Hamiltonian diagonalisation}\label{sec:classical-hd}

To obtain a quantum theory from the semiclassical action (\ref{eq:action-mukhanov}), the field $v$ is promoted to an operator: 
\be\label{eq:operator-v} 
\hat{v}(\eta,x) = \int
    \frac{\mathop{d^3k}}{(2\pi)^3}\Big[ \hat{a}_k v_k(\eta) e^{ik\cdot x} + \hat{a}^{\dagger}_k
    v^{\ast}_k(\eta) e^{-ik\cdot x} \Big],
\ee
where $\hat{a}_k^{\dagger}$ and $\hat{a}_k$ are the creation and annihilation operators, respectively.  The momentum conjugate to $v$ is 
\be\label{eq:v-conj-mom} \pi = \frac{\partial \mathcal{L}}{\partial v'} = v', \ee
(with $\mathcal{L}$ being the Lagrangian density associated with the action) and is quantised accordingly. To impose canonical commutation relations
\be\label{eq:canonical-comm-a} \left[ \hat{a}_k, \hat{a}_{k'}^{\dagger} \right] = (2\pi)^3
    \delta(k - k'), \ee
and to obey quantum dynamics, the time-dependent part of the mode functions in (\ref{eq:operator-v}) must satisfy the equation of motion and a normalisation constraint: 
\begin{align}
v_k'' + \left( k^2 - \frac{z''}{z} \right)v_k &= 0, \label{eq:eom-op-v}\\
v_k'v_k^{\ast} - v_k^{\ast}{}'v_k &= -i.\label{eq:norm-v}
\end{align}
The equations (\ref{eq:eom-op-v})--(\ref{eq:norm-v}) do not fully determine the mode functions $v_k(\eta)$. To fix the leftover degree of freedom, one needs to specify a vacuum, which obeys $\hat{a}_k\ket{0} = 0$.  One popular definition of the ground state is that which minimises the vacuum expectation value of the Hamiltonian of the system.  The choice of variables $(\eta, v)$ results in the expression
\be\label{eq:hamiltonian-vev-v} 
\bra{0} H \ket{0} \propto \int \frac{\mathop{d^3k}}{(2\pi)^3}\Bigg[ |v_k'|^2 + \left( k^2 - \frac{z''}{z} \right)^2 |v_k|^2\Bigg].
\ee
We must then minimise the contribution to the Hamiltonian separately for each $k$-mode, with respect to the mode functions belonging to that mode, subject to the constraint (\ref{eq:norm-v}). This leads to the solutions
\be\label{eq:ic-hd-v} 
\begin{split} |v_k|^2 &= \frac{1}{2\sqrt{ k^2 - \frac{z''}{z}}}, \\ v_k'
    &= -i\sqrt{k^2 - \frac{z''}{z}}v_k, \end{split} \ee
which can be used to set initial conditions on the mode functions $v_k$.

\subsubsection{Danielsson vacuum}\label{sec:classical-danielsson}

Danielsson \cite{danielsson} proposed that the vacuum should instead be chosen such that
\be\label{eq:danielsson-ic-v} 
\begin{split} |v_k|^2 &= (2k)^{-1}, \\ v_k' &= \left( -ik +
\frac{a'}{a} \right)v_k.  \end{split} \ee
This result is derived from `first principles', working in the Heisenberg picture.  For detailed reviews of the following, see \cite{easther,polarski-starobinsky,alberghi}.

In the Heisenberg picture, the operators carry time-dependence.  When quantising the massless field $y$ appearing in the action (\ref{eq:action-scalar}) with $m=0$, we may write in Fourier space 
\be
\hat{y}_k(\eta) = \frac{1}{\sqrt{2k}}\left[\hat{a}_k(\eta) + \hat{a}_{-k}^{\dagger}(\eta)\right],
\ee
and for its conjugate momentum $\pi_k = y_k' - \frac{a'}{a}y_k$, 
\be \hat{\pi}_k(\eta) =
-i\sqrt{\frac{k}{2}}\left[\hat{a}_k(\eta) - \hat{a}_{-k}^{\dagger}(\eta)\right].  \ee
The creation and annihilation operators mix over time via a Boguliubov transformation
\be 
\begin{split}
    \hat{a}_k(\eta) &= \alpha_k(\eta)\hat{a}_k(\eta_0) + \beta_k(\eta)\hat{a}_{-k}^{\dagger}(\eta_0),\\
\hat{a}_{-k}^{\dagger}(\eta) &= \beta_k^{\ast}(\eta)\hat{a}_k(\eta_0) +
\alpha_k^{\ast}(\eta)\hat{a}_{-k}^{\dagger}(\eta_0),  \end{split} \ee
where $\alpha_k(\eta)$ and $\beta_k(\eta)$ are time-dependent mixing coefficients.
One can therefore isolate the time-dependent parts of the fields, 
\be 
\begin{split} \hat{y}_k(\eta) &= f_k(\eta)\hat{a}_k(\eta_0)
+ f_k^{\ast}(\eta)\hat{a}_{-k}^{\dagger}(\eta_0),\\ i\hat{\pi}_k(\eta) &= g_k(\eta)\hat{a}_k(\eta_0)
- g_k^{\ast}(\eta)\hat{a}_{-k}^{\dagger}(\eta_0), \end{split} \ee
 with 
\be 
\begin{split} f_k(\eta)
    &= \frac{1}{\sqrt{2k}}(\alpha_k(\eta) + \beta_k^{\ast}(\eta)),\\ g_k(\eta) &=
\sqrt{\frac{k}{2}}(\alpha_k(\eta) - \beta_k^{\ast}(\eta)), \end{split} \ee
where the $f_k$ now take the role of the mode functions $y_k$ in the Schr\"{o}dinger picture, because they carry all time dependence of the field operator. At $\eta = \eta_0$, the creation and annihilation operators are by definition unmixed, therefore 
\be \beta_k(\eta_0) = 0 = \sqrt{\frac{k}{2}}f_k^{\ast}(\eta_0) -
\frac{1}{\sqrt{2k}}g_k^{\ast}(\eta_0).  \ee
 Identifying the mode function $y_k$ with $f_k$ and the
conjugate momentum $\pi_k$ with $-ig_k$, we obtain the Danielsson vacuum (\ref{eq:danielsson-ic-v}).

Immediately it is clear that since the Danielsson vacuum relates a field and its conjugate momentum, the initial conditions derived from it will generally change under transformations that change that relationship, e.g. the addition of a surface term to the action, described in \cref{sec:meth-surf-terms}.

\subsubsection{Minimising the renormalised stress--energy tensor}\label{sec:classical-rst} 

\citet{nqicfi} have proposed to determine the ground state by minimising the vacuum expectation value of the local energy density to avoid the excessive particle production of the Hamiltonian diagonalisation approach.  The local energy density is computed as the $00$ component of the stress--energy tensor of the system (\ref{eq:action-scalar}). General relativity requires a symmetric stress--energy tensor, as it appears on the right-hand side of the Einstein equations. It is defined as \cite{birrell-davies}
\be\label{eq:gr-T}
T_{\mu \nu} = -\frac{2}{\sqrt{-g}}\frac{\delta \mathcal{S}_\mathrm{m}}{\delta g^{\mu \nu}}, 
\ee
with the subscript m signalling the matter part of the action. 
Just like the expectation value of the Hamiltonian, the stress--energy tensor is divergent, and has to be renormalised to yield finite quantities.  There exist several procedures of renormalisation of the stress--energy tensor, and in \citet{nqicfi} the Hadamard point-splitting method is used (thoroughly described in \cite{birrell-davies}).  In summary, this consists of first quantising the field $y$ from \cref{sec:scalar-fields}, writing down an expression for the Hadamard Green function,
\be\label{eq:hadamard-g1}
G^{(1)}(x, x') = \tfrac{1}{2}\bra{0}\{ \varphi(x), \varphi(x')
\}\ket{0}
\ee
then applying the bi-scalar derivative $\mathcal{D}_{\mu \nu}$ to 
\be G^{(1)}(x, x') - G^{(1)}_{\text{DS}}(x, x'),
\ee
where the second term denotes the de-Witt Schwinger geometrical terms. The geometrical terms do not depend on the variables with respect to which one minimises the stress--energy tensor, and are therefore summarised as $\tilde{T}$ and ignored in the minimisation process.  The coincidence limit $x \rightarrow x'$ is then taken to yield $\bra{0} T_{\mu \nu} \ket{0}_{\text{ren}}$. Altogether,
\be \label{eq:rst}
\begin{gathered}
\bra{0}T_{\mu \nu}\ket{0}_{\text{ren}} = \lim_{x\to x'}\tfrac{1}{2} \Big[
\big(\nabla_{\mu}\nabla_{\nu'} + \nabla_{\mu'}\nabla_{\nu}\big) \\ - g_{\mu
\nu}\nabla_{\alpha}\nabla^{\alpha'} + g_{\mu \nu} m^2 \Big]G^{(1)}(x,x') + \tilde{T}, 
\end{gathered}
\ee
which is a functional to be minimised with respect to the mode functions $\{y_k, y_k^{\ast},y_k', y_k^{\ast}{}'\}$ treated as independent variables, subject to the normalisation constraint on $y_k$ arising from the canonical commutation relations.  The minimisation process yields the solutions  
\be\label{eq:rst-ic-y} 
\begin{gathered}
|y_k|^2 = \frac{1}{2\sqrt{\left(k^2 + m^2a^2\right)}},\\
y_k' = \left( -i\sqrt{k^2 + m^2a^2} + \frac{a'}{a}\right)y_k, 
\end{gathered}
\ee
which, under the circumstances described in \cref{sec:scalar-fields}, can be used to set
\be\label{eq:rst-ic-v} 
\begin{split}
|v_k|^2 &= (2k)^{-1}, \\
v_k' &= \left( -ik + \frac{z'}{z} \right)v_k 
\end{split} 
\ee
for the Mukhanov variable.

\section{Methods}\label{sec:methods}

In the theories of primordial perturbations reviewed in the previous section, the standard choices of variables was explained by the simple form of equation of motion they admitted, more specifically that the modes involved behaved like harmonic oscillators, so their quantisation was known. 

However, as we shall show, one can always make a canonical transformation of the action of the system in the extended phase space by redefining the independent and dependent variables simultaneously such that the resulting equation of motion is first-derivative-free and appropriate commutation relations are satisfied. 
The new, redefined field will then still be quantisable by analogy with the harmonic oscillator. Apart from the conveniently `bare' $k^2$ term in the associated equation of motion (originating from the canonically normalised scalar field in the action), there is nothing that makes the standard choice of variables $(\eta, v)$ special. In fact there is no choice of variables for which the action is canonically normalised in the case of a non-flat universe \cite{handley-curved}.
In addition to field-redefinitions, one could always add a surface term to the Lagrangian, equivalent to performing an integration by parts, and obtain a dynamically equivalent system (so long as the boundary terms vanish). 
This section summarises the procedures described above by which the vacuum prescriptions will be transformed.

\subsection{Field redefinition}\label{sec:meth-field-redef}

For all vacuum-setting methods involving the quantisation of a field, we consider quantising an alternative field related to the conventional choice by a time-dependent, scalar-valued, homogeneous function $h$, and the redefinition of time to a new independent variable $\tau$, whilst keeping in mind how the form of the metric changes. We then derive the vacuum conditions in analogy with the conventional procedures. For Hamiltonian diagonalisation this redefinition will thus be 
\be\label{eq:field-redef-hd}
t \to \tau(t), \quad \mathcal{R} \to \chi(x, \tau) \equiv \frac{\mathcal{R}}{h(\tau)},
\ee
and for minimising the renormalised stress--energy tensor, we shall consider instead
\be\label{eq:field-redef-rst}
t \rightarrow \tau(t), \quad \varphi \rightarrow \chi(x, \tau) \equiv \frac{\varphi}{h} = \frac{y}{ah},
\ee
as the field being quantised is $\varphi$.

We shall derive a constraint linking $\tau$ and $h$ for each action considered to ensure it yields the equation of motion of an oscillator, as this is not guaranteed by the transformation (\ref{eq:field-redef-hd}). This generally leaves an unconstrained integration constant $C_0$. The field-redefinition also does not necessarily conserve the commutator structure. For all transformations considered, this will be ensured via an additional constraint that needs to be satisfied during the minimisation of the vacuum expectation value of the Hamiltonian density or the $00$-component of the renormalised stress--energy tensor.

\subsection{Surface terms}\label{sec:meth-surf-terms}

In the conventional Hamiltonian diagonalisation approach, one performs an integration by parts to obtain the action (\ref{eq:action-mukhanov}). This is equivalent to adding the total derivative
\be\label{eq:mukhanov-boundary-term}
\left( \frac{z'}{z}v^2 \right)'
\ee
to the associated Lagrangian density, or a vanishing \emph{surface or boundary term} to the action.
Under the field redefinition of the previous section, the action changes such that the appropriate boundary term to add will be
\be\label{eq:gen-boundary-term}
-\partial_{\tau}\left( \chi^2 \frac{\partial_{\tau} h}{h} \right). 
\ee
It is clear that (i) there are infinitely many choices of vanishing boundary terms one could add; and that (ii) the boundary terms modify the form of the action and in turn the field-conjugate momentum relationship. For Hamiltonian diagonalisation and the Danielsson vacuum we shall investigate how the addition of the boundary term (\ref{eq:gen-boundary-term}) alters the initial conditions, and we shall show that minimising the renormalised stress--energy tensor is invariant under the addition of boundary terms by construction.  

Detailed calculations of initial conditions arising from systems subject to field-redefinitions and addition of surface terms can be found in \cref{sec:hd-gen,sec:rst-gen}. 

\section{Results}\label{sec:results}

\subsection{Initial conditions}\label{sec:results-ic}

\subsubsection{Hamiltonian diagonalisation}\label{sec:hd-ic}

Under the field redefinition (\ref{eq:field-redef-hd}), minimising the vacuum expectation value of the Hamiltonian (\ref{eq:hamiltonian-vev-v}) gives the generalised initial conditions 
\be\label{eq:ic-hd-chi} 
\begin{split} |\chi_k|^2 &= \left(2C_0\omega_k\right)^{-1},\\
\partial_{\tau}\chi_k &= -i\omega_k\chi_k,  \end{split} \ee
where $\omega_k$ is the time-dependent frequency of the equation of motion in terms of $(\tau, \chi_k)$, and $C_0$ is a constant.  

As a sanity check, substituting $h=z^{-1}$ and $C_0=1$, corresponding to quantising the Mukhanov variable in terms of conformal time, we recover the conventional initial conditions (\ref{eq:ic-hd-v}). However, changing variables to $(\eta, v)$ in the generalised initial conditions (\ref{eq:ic-hd-chi}) yields
\be\label{eq:ic-hd-v-gen} 
\begin{gathered}
|v_k|^2 = \frac{1}{2\sqrt{k^2 + \frac{h''}{h} + 2\left( \frac{h'z'}{hz} \right)}}, \\ 
v_k' =\Bigg( -i\sqrt{ k^2 + \frac{h''}{h} + 2\frac{h'z'}{hz}} 
+ \frac{h'}{h} + \frac{z'}{z} \Bigg) v_k, 
\end{gathered}
\ee
which carry the arbitrary function $h$.  This means that one can derive a family of initial conditions for $v_k(\eta)$ depending on the field in terms of which the action was written.  

Moreover, under addition of the boundary term defined in \cref{sec:meth-surf-terms} and the field re-definition that led to (\ref{eq:ic-hd-v-gen}), one can derive another set of conditions:
\be\label{eq:hd-ic-v-gen-bt} 
\begin{gathered}
|v_k|^2 = \frac{1}{2\sqrt{k^2 - \left(\frac{h'}{h}\right)^2}},\\ v_k' = \Bigg( -i\sqrt{ k^2 - \left(\frac{h'}{h}\right)^2 }
+ \frac{h'}{h} + \frac{z'}{z} \Bigg)v_k.  
\end{gathered}
\ee
Not only is this another family of solutions depending on the function $h$, it is a different set to (\ref{eq:ic-hd-v-gen})! 
This can be seen by noting that for all values of $h'/h$ in (\ref{eq:hd-ic-v-gen-bt}) there will be a lower limit for $k$ below which the expression under the square root becomes negative, which is not allowed (for the squared magnitude of the perturbation has to be real). This `forbidden region' is different for (\ref{eq:ic-hd-v-gen}).
Since we could have chosen any action connected to (\ref{eq:action-mukhanov}) via a canonical transformation, we could have arrived at a range of different Hamiltonians yielding different quantum initial conditions for the perturbations.  Hamiltonian diagonalisation thus gives ambiguous initial conditions depending on the choice of canonical variables. 

The two example solution families obtained via Hamiltonian diagonalisation are parametrised by a time-dependent function $h$. 
If we choose to use the solutions as initial conditions, and the spacetime slice we set them on is chosen such that all modes (i.e.\ all $k$) are set simultaneously at a conformal time $\eta_0$, the shape of the function $h(\eta)$ only matters in the vicinity of $\eta_0$.  We can determine how much freedom this gives by Taylor-expanding $h$ near $\eta_0$, 
\be\label{eq:taylor-h} 
h(\eta) = \sum_{n=0}^{\infty}\frac{1}{n!}h_n(\eta - \eta_0)^n , 
\ee
substituting into (\ref{eq:ic-hd-v-gen}), then finally evaluating at $\eta = \eta_0$. This gives
\be\label{eq:ic-hd-v-gen-dof1}
\begin{gathered}
|v_k|^2 = \frac{1}{2\sqrt{k^2 + \frac{h_2}{h_0} + 2\left( \frac{h_1z'}{h_0z} \right)}}, \\ 
v_k' = \Bigg(-i\sqrt{ k^2 + \frac{h_2}{h_0} + 2\frac{h_1z'}{h_0z}} + \frac{h_1}{h_0} + \frac{z'}{z} \Bigg)v_k ,
\end{gathered}
\ee
showing that there are two real degrees of freedom in this set of initial conditions ($h_2/h_0$ and $h_1/h_0$), whereas (\ref{eq:hd-ic-v-gen-bt}) only depends on $h_1/h_0$: 
\be\label{eq:ic-hd-v-gen-dof2}
\begin{gathered}
|v_k|^2 = \frac{1}{2\sqrt{k^2 - 2\left( \frac{h_1}{h_0} \right)^2}}, \\ 
v_k' = \Bigg( -i\sqrt{ k^2 - 2\left(\frac{h_1}{h_0}\right)^2 } + \frac{h_1}{h_0} + \frac{z'}{z} \Bigg)v_k.
\end{gathered}
\ee
It is worth noting that there exist prescriptions that do not set initial conditions for all perturbations simultaneously: \cite{danielsson} for example chooses a finite $\eta_0(k)$ for each $k$ such that the physical momentum corresponding to the mode is given by some fixed scale, and it is common to initialise the perturbations when their lengthscales are a fixed fraction of the Hubble horizon. These prescriptions may lead to more degrees of freedom in the initial conditions, but we leave such considerations to future work.

In \cite{Lim}, the authors do not attempt to find the preferred choice of vacuum (if that exists), rather they parametrise a general choice of initial conditions by two complex scalars, $X$ and $Y$ ($Y$ being purely imaginary), that characterise the choice of vacuum: 
\begin{gather}\label{eq:lim-ic} 
v_k(\eta_0) = \frac{e^{i\phi_1}}{\sqrt{k}}\left[ 1 + \frac{X+Y}{2}\theta_0 + \mathcal{O}(\theta_0^2)\right], \\ 
v_k'(\eta_0) = -i\sqrt{k}e^{i\phi_2}\left[ 1 + \frac{Y-X}{2}\theta_0 + \mathcal{O}(\theta_0^2) \right].
\end{gather} Their initial conditions are the solutions of the mode equation (\ref{eq:ms-v}) in the limit where the metric is Minkowski, with added corrections due to expansion as a power series in the small, dimensionless parameter $\theta_0$:
\be\label{eq:theta-0}
\theta_0 = \frac{aH}{k}\Big|_{\eta = \eta_0}. 
\ee
In terms of this parametrisation, the solutions (\ref{eq:ic-hd-v-gen-dof1}) represent a subset with
\be\label{eq:lim-param}
X = -Y, \quad Y \;\text{arbitrary}, 
\ee
which can be seen by factoring out $1/\sqrt{k}$ from the expression for $v_k$, expanding the rest in powers of $1/k$, and observing that there is no term linear in $1/k$.

\subsubsection{Danielsson vacuum}\label{sec:danielsson-ic}

It is easily seen that Danielsson's result, for a different choice of field (\ref{eq:field-redef-rst}), generalises as  
\be
\pi_{\chi}(\tau) = -ik\chi(\tau),
\ee
where $\pi_{\chi}$ is the momentum conjugate to $\chi$.  Using the action (\ref{eq:action-scalar}) under the field redefinition (\ref{eq:field-redef-rst}) to compute the conjugate momentum, this gives for $y_k$: 
\be\label{eq:dani-ic-1} 
\begin{gathered}
|y_k|^2 = \frac{(ah)^2}{2k}, \\ 
y_k' = \left( -\frac{ik}{(ah)^2} + \frac{a'}{a} \right)y_k.
\end{gathered}
\ee
Under the addition of the surface term (\ref{eq:gen-boundary-term}) to the action (which has the effect of eliminating the $\partial_{\tau}\chi$ term), the momentum conjugate to the field $\chi$ changes, and the initial conditions become
\be\label{eq:dani-ic-2}
\begin{gathered}
|y_k|^2 = \frac{(ah)^2}{2k}, \\
y_k' = \left( -\frac{ik}{(ah)^2} + \frac{a'}{a} + \frac{h'}{h} \right)y_k.  
\end{gathered} 
\ee
Since canonical transformations by addition of surface terms only change the relationship between the field and its conjugate momentum, infinitely many solution families like (\ref{eq:dani-ic-1}) and (\ref{eq:dani-ic-2}) exist that differ in $y_k'(y_k)$.  There is no theoretical guidance as to which action one should consider, and hence which set of initial conditions is the correct one in the Danielsson prescription.

\subsubsection{Minimising the renormalised stress--energy tensor}\label{sec:rst-ic}

Working from the action (\ref{eq:action-scalar}) under a canonical field redefinition, and minimising the expression (\ref{eq:rst}) fixes the mode functions $\chi_k$ fully as 
\be\label{eq:rst-ic-chi} 
\begin{gathered}
|\chi_k|^2=\frac{1}{2\sqrt{h^4a^4\left(k^2 + m^2a^2\right)}},\\ 
\partial_{\tau}\chi_k = \Bigg( -\frac{i}{C_0}\sqrt{h^4a^4\left(k^2 + m^2a^2\right)}
-\frac{\partial_{\tau}h}{h}\Bigg)\chi_k.  
\end{gathered}
\ee
Converting back to conformal time and $y$ gives the solutions 
\be\label{eq:rst-ic-y-gen} 
\begin{gathered}
|y_k|^2 = \frac{1}{2\sqrt{\left(k^2 + m^2a^2\right)}},\\ y_k' = \left( -i\sqrt{k^2 +
m^2a^2} + \frac{a'}{a}\right)y_k, 
\end{gathered}
\ee
independently of the choice of $h(\tau)$ and $\tau$, and in agreement with solutions (\ref{eq:rst-ic-y}) derived using the conventional treatment. 

From the definition (\ref{eq:gr-T}), it can be shown that an added 4-divergence (involving the field and the metric) to the Lagrangian does not change the form of the stress--energy tensor as long as the added surface term does not break covariance of the existing Lagrangian. Not only has the stress--energy tensor been constructed in a covariant way, \citet{yargic} have recently shown that one can define a ground state family, with respect to which its expectation value is covariantly conserved. This possibility was also pointed out in \citet{Fulling1979}. All that matters in the minimisation process is the form of the field-dependent part of $T_{\mu \nu}$ (the geometrical terms do not come into play as they do not depend on the field), and therefore the initial conditions derived via minimisation of the renormalised stress--energy tensor are invariant under the addition of surface terms.

Minimising the $00$-component of the renormalised stress--energy tensor is therefore invariant under canonical field-redefinitions and addition of surface terms. The stress--energy tensor was defined to be covariant, therefore its invariance under the addition of surface terms comes as no surprise, but it has not been constructed to be invariant under field-redefinitions. 
In taking the $00$-component, however, covariance has been broken, and we must ensure that the same results hold when a coordinate-independent quantity is minimised. 

To make sure this is the case, one should consider the eigenvalue belonging to a timelike eigenvector of the tensor. Let us therefore generate the off-diagonal components of $\bra{0} T_{\mu \nu}\ket{0} $, ignoring those arising from the geometrical terms: 
\be\label{eq:i0} 
\bra{0}T_{i0}\ket{0}_{\text{ren}} = \tfrac{1}{2} \int \frac{d^3k}{(2\pi)^3} \frac{k_i}{\dot{\tau}az^2} , 
\ee
\be\label{eq:ij}
\bra{0}T_{ij}\ket{0}_{\text{ren}} = \tfrac{1}{2} \int \frac{d^3k}{(2\pi)^3} 2 h^2 \chi_k
\chi_k^{\ast} k_i k_j ,
\ee
with $i\neq j$.  All the above components include integrals with integrands odd in $k_i$, and so they vanish.  The de Witt--Schwinger terms in the off-diagonal elements can also be shown to vanish due to the metric $g_{\mu \nu}$ being diagonal \footnote{For an explicit expression for $\bra{0}T_{\mu \nu}\ket{0}_{\text{ren}}$ that confirms this, see \cite{birrell-davies}.}.  Consequently the expectation value of the renormalised stress--energy tensor is diagonal, and its eigenvalues are just the diagonal entries.  The eigenvalue corresponding to the timelike eigenvector is the $00$ element, confirming the results of this section.

\subsubsection{Generalised analogy between a scalar field on a curved spacetime and the curvature perturbation}\label{sec:equivalence}

In \cref{sec:classical-rst}, we derived initial conditions for the Mukhanov variable in an inflating background from the solutions for $y_k$ in (\ref{eq:rst-ic-y}) by drawing an analogy between their equations of motion, (\ref{eq:ms-v}) and (\ref{eq:eom-scalar}). For the two equations to be the same, we required $a''/a = z''/z$.
We now derive similar requirements for the two equations of motion to have the same form, after having performed the field-redefinitions
\be
\mathcal{R} \rightarrow \chi_1 = \frac{\mathcal{R}}{h_1}
\ee 
and
\be
\varphi \rightarrow \chi_2 = \frac{\varphi}{h_2} = \frac{y}{ah_2},
\ee
respectively. The first, obvious requirement is that the redefined field in (\ref{eq:eom-scalar}) has to be massless, $m=0$.  For the classical perturbed action we then have the equation of motion
\be 
0 = \partial_{\tau_1\tau_1} \chi_1 + \Bigg[ \bigg(\frac{kh_1^2z^2}{C_1^2}\bigg)^2 + \frac{ \partial_{\tau_1\tau_1} h_1}{h_1} - 2\bigg(\frac{\partial_{\tau_1}h_1}{h_1}\bigg)^2 \Bigg]\chi_1,
\ee
whereas from the massless scalar field we get 
\be
0 = \partial_{\tau_2\tau_2} \chi_2 + \Bigg[ \bigg(\frac{kh_2^2a^2}{C_2^2}\bigg)^2 + \frac{ \partial_{\tau_2\tau_2} h_2}{h_2} - 2\bigg(\frac{\partial_{\tau_2}h_2}{h_2}\bigg)^2 \Bigg]\chi_2.  
\ee
We are free to choose $h_1$, $h_2$, and the constants $C_1$ and $C_2$. We first have to set 
\be\label{eq:id-eom-1}
\frac{h_1z}{C_1} = \frac{h_2a}{C_2}
\ee
for the first term in the frequency to be the same, then we need to satisfy 
\be\label{eq:id-eom-2}
\frac{h_1''}{h_1} - 2\left(\frac{h_1'}{h_1}\right)^2 = \frac{h_2''}{h_2} - 2\left(\frac{h_2'}{h_2}\right)^2, 
\ee
where prime denotes differentiation with respect to the function's argument. This differential equation admits the solution 
\be\label{eq:id-eom-2-sol} 
\frac{h_1}{h_2} = A\exp\left({\int \mathop{dt} \frac{h_2^2 }{B - \int \mathop{dt} h_2^2 }}\right).  
\ee
One can always choose $h_1/h_2$ such that (\ref{eq:id-eom-1}) holds. One can then calculate $h_2$ by inverting (\ref{eq:id-eom-2-sol}).  This means that if $h_1$, $h_2$ are chosen appropriately, one can always map the solution of the scalar-field action onto the cosmological perturbations. 

For an arbitrary choice of $h_1$ and $h_2$, if $z \propto a$, the equations of motion are guaranteed to take the same form in general, therefore the mapping from the test scalar field to the primordial perturbations becomes the identity. (Note that for a specific choice of $h_1$ and $h_2$, e.g.\ $h_1 = 1/z$, $h_2=1$, the requirement for form-similarity may be less restrictive.)

Overall, the conclusion that we may use the solutions (\ref{eq:rst-ic-chi}) as initial conditions for the associated perturbation mode in inflation is still valid for a different choice of (extended) canonical variables.

\subsubsection{Other vacuum choices}\label{sec:other-vacua}

\paragraph{Bunch--Davies vacuum:}
A well-known vacuum prescription left out of our analysis is the Bunch--Davies vacuum \cite{bunch-davies-vacuum}, which is set by requiring that the positive-frequency mode function matches the Minkowski solution in the limit of all mode lengthscales being subhorizon ($k/(aH) \rightarrow \infty$), in the infinite conformal past. The prescription only specifies the vacuum uniquely if inflation did not have a start, and otherwise results in ambiguities associated with the artificially imposed $k\rightarrow \infty$ limit \cite{chung-notari-riotto}. This issue aside, the Bunch--Davies prescription relies on the choice of canonical variables being $(\eta, v_k)$, because it yields an oscillator equation of motion with a `bare' $k^2$ term in the squared frequency, thus the frequency can be adiabatically constant. This is no longer the case when the equation of motion is rewritten in terms of a different time and field, i.e.\ under field-redefinitions. This preference for variable choice is the reason why we do not consider the Bunch--Davies vacuum.

\paragraph{Adiabatic vacuum:}
The adiabatic vacuum formalism \cite{birrell-davies,chung-notari-riotto} is similar to the Bunch--Davies one in that it states that sometimes a unique vacuum can be defined when the frequency of the perturbation modes is adiabatically constant, and given by an $n^{\mathrm{th}}$ order Wentzel--Kramers--Brillouin (WKB) approximation. Since canonical transformations leave the equation of motion invariant, they will not affect the adiabatic vacuum if the expansion is carried out to arbitrarily high order, but otherwise the (truncated) WKB solution of an equation of motion rewritten in terms of different independent/dependent variables will generally differ. This is easily seen by observing that the lowest order adiabatic initial conditions match those obtained via Hamiltonian diagonalisation (\ref{eq:ic-hd-chi-appendix}).

\paragraph{$\alpha$-vacua:}
The behaviour of $\alpha$-vacua \cite{goldstein-lowe-alphavacua} under canonical transformations has been reviewed in \cite{venningrain2019}. These vacua were found not to be canonically invariant.

\subsection{Primordial power spectra}\label{sec:results-pps}

\begin{figure}[tbp]
	\centering
    \includegraphics[width=\columnwidth]{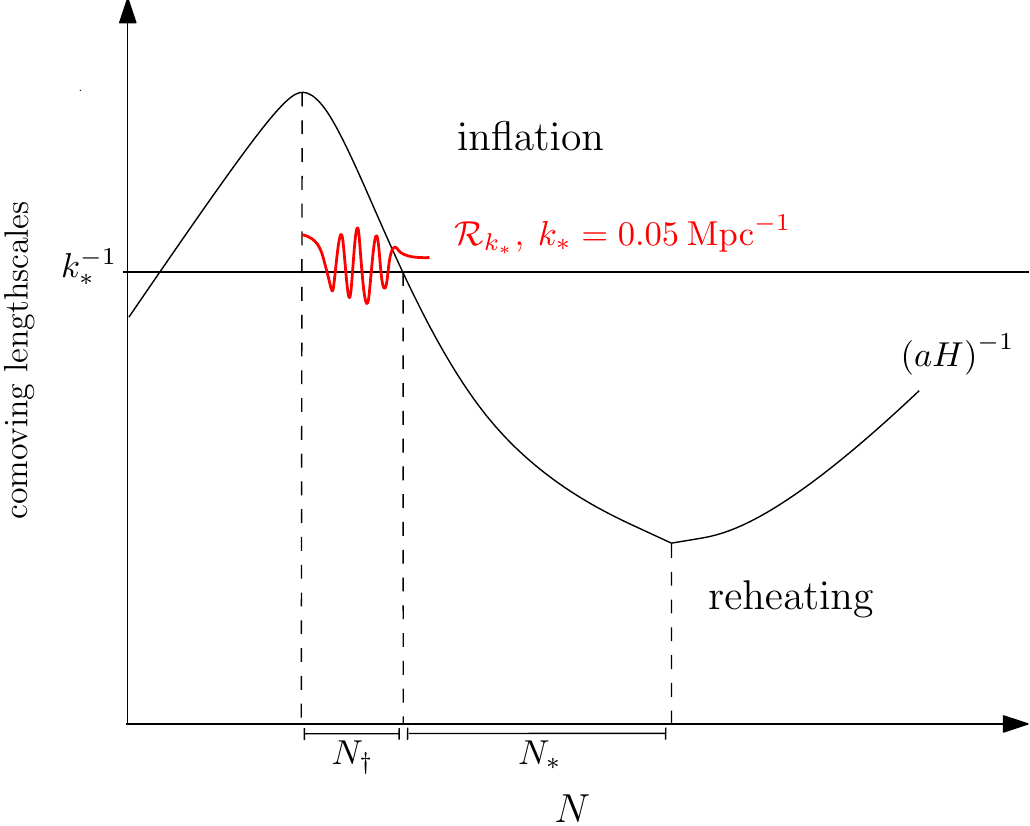}    \caption{\label{fig:evolution} Schematic evolution of the comoving Hubble horizon (black) and a `pivot' perturbation $\mathcal{R}_{k_\ast}$ (red), whose characteristic lengthscale is \SI{0.05}{\per\mega\parsec}. Inflation starts as the comoving horizon begins to shrink, and continues until it reaches its minimum. The total number of e-folds by which the universe has grown during inflation is $N_{\mathrm{tot}} = N_{\ast} + N_{\dagger}$, with $N_{\ast}$ e-folds passing between the pivot perturbation exiting the horizon and the end of inflation. }
\end{figure}

Primordial power spectra show the power in the Fourier modes of gauge-invariant curvature perturbations ($\mathcal{R}_k$) after horizon exit -- when their characteristic lengthscale $k^{-1}$ first exceeds the size of the comoving Hubble horizon $(aH)^{-1}$ during inflation. When this happens, the amplitude of fluctuations becomes constant, as shown in \cref{fig:evolution} alongside definitions of quantities used in this section. 
Primordial power spectra are not directly observable\footnote{They are, however, reconstructable \cite{handley-reconstruction,Planck2015infl,Planck2018infl}.}, but are the first physical quantities one can derive from the perturbations, and hence the first physical objects affected by the existence of a range of quantum initial conditions. To compute them, first the `cosmological background' quantities such as the scale factor and Hubble horizon are computed by solving the cosmological field equations \cite{friedmann}, which are derived from the action (\ref{eq:classical-action}) using Einstein's field equations. 

In this work, it is assumed that slow-roll inflation had a start and was preceded by a non-inflationary phase. An eternal inflation scenario would allow one to initialise all primordial perturbation modes deep inside the Hubble horizon, which would result in a power-law primordial power spectrum that is independent of the choice of primordial initial conditions and the choice of vacuum. Our result concerning the (non-)invariance of vacuum choices under field-redefinitions and addition of surface terms would be theoretically important nonetheless, but the aim of this and the following sections is to demonstrate the observational impact of the choice of canonical variables, and therefore we choose a physically well-motivated setup in which the choice of vacuum affects the primordial power spectrum. Generally it may also be attractive to choose a model without too long an inflationary phase in order to avoid having to account for trans-Planckian physics when initialising the primordial perturbations (see \cite{martin-brandenberger-transplanckian-1,martin-brandenberger-transplanckian-2, Lim, danielsson} and references therein), due to lengthscales observed today having been smaller than the Planck length at some earlier epoch.

The numerical solution is initiated at the start of inflation, which is assumed to be preceded by a \emph{kinetically dominated} \cite{kineticic,hergt-1,hergt-2} phase.  
In kinetic dominance the inflaton particle's kinetic energy dominates over its potential energy, causing the comoving Hubble horizon to grow and reach a maximum as the universe enters slow-roll inflation.

For simplicity, we consider a single-field inflationary model with a potential that is quadratic in the field,
\be
V(\phi) = \tfrac{1}{2}\mu^2\phi^2.
\ee

For convenience, we choose the number of e-foldings, $N \equiv \ln{a}$ to be the independent variable in the field equations. With this choice, they become
\begin{align}
    \frac{d\ln{D}}{dN} &= 4 + D\left(4K-2e^{2N}V(\phi)\right) \label{eq:field-eq-1}, \\
    \left( \frac{d\phi}{dN} \right)^2 &= 6 + D\left( 6K - 2e^{2N}V(\phi) \right) \label{eq:field-eq-2},
\end{align}
where $D = (aH)^{-2}$ and $K$ is the spatial curvature that can take values $0$, $\pm 1$ for flat, closed, and open universes, respectively \cite{agocs}. 
We set $K=0$ when calculating the primordial power spectra and all subsequent computations. 
At the start of inflation, the field equations give 
\begin{align}
   D = D_i = \frac{2 e^{-2N_i}}{V(\phi_i)},
\end{align}
leaving $N_i$ and $\phi_i$ as adjustable parameters.
In a flat universe we can set $N_i=0$ without loss of generality, so the only free parameter left in the background initial conditions is the initial field strength, $\phi_i$. This determines $N_{\mathrm{tot}}$, the total number of e-folds during inflation, and $N_{\dagger}$, the number of e-folds between the start of inflation and a `pivot' perturbation mode with lengthscale $k_{\ast}^{-1} =$ \SI{0.05}{\per\mega\parsec} exiting the Hubble horizon.
Observations constrain the value of $N_{\ast}$, the number of e-folds between the pivot scale exiting the Hubble horizon and the end of inflation, to be between 50 and 60 \cite{Liddle1993,Liddle2003,Dodelson2003}.
We pick a value for $\phi_i$ consistent with the `just enough inflation' scenario \cite{Ramirez2012-1, Ramirez2012-2}, where $N_{\ast} \sim N_{\mathrm{tot}}$. 
In the inflationary potential, $\mu$ denotes the inflaton mass, which determines the overall `normalisation' of the primordial power spectrum and can be calculated using the slow-roll approximation from $N_{\ast}$ and $A_s$ to yield the observed normalisation.
The primordial parameters and the cosmological parameters used in the following sections are summarised in \cref{tab:cmbparams}.

Once the cosmological field equations have been solved, the perturbations $\mathcal{R}_k$ are initialised simultaneously (independently of $k$) at the start of inflation with two arbitrary sets of initial conditions (e.g.\ $\{\mathcal{R}_k = 1, d\mathcal{R}_k/dN =0\},\; \{\mathcal{R}_k = 0, d\mathcal{R}_k/dN = 1\}$) and for each $k$-mode the Mukhanov--Sasaki equation is solved to obtain their evolution in $N$. We use the numerical solver \texttt{oscode} \cite{agocs} to carry out this computation efficiently. The amplitude of each perturbation is read off when they are well outside the Hubble horizon, $k < 10^{-2}aH$. The two solutions for $\mathcal{R}_k$ can then be linearly combined to satisfy any initial condition at the start of inflation from the sets derived, without having to re-compute the evolution of perturbations.
Finally, the primordial power spectrum is constructed according to
\begin{equation}
    \mathcal{P}_{\mathcal{R}}(k) = \frac{k^3}{2\pi^2}|\mathcal{R}_k|^2.
\end{equation}

\begin{figure}[tbp]
	\centering
    \includegraphics[scale=0.99]{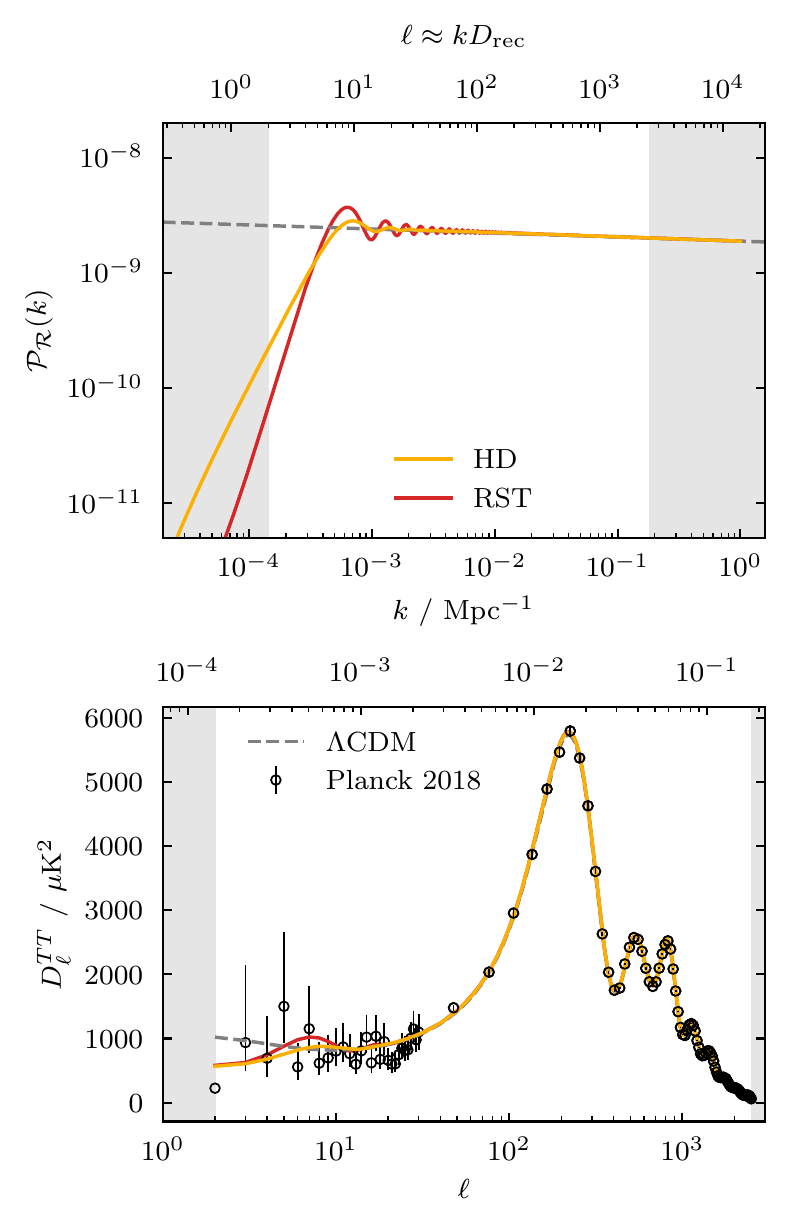}
    \caption{\label{fig:ppscmbhdrst}Primordial power spectrum (top panel) and angular CMB power spectrum of temperature anisotropies (bottom panel) for the standard Hamiltonian Diagonalisation~(HD) and Renormalised Stress-Energy-Tensor~(RST) vacuum conditions.}
\end{figure}

\cref{fig:ppscmbhdrst} shows example primordial power spectra generated using conventional Hamiltonian diagonalisation (\ref{eq:ic-hd-v}) and renormalised stress--energy tensor (\ref{eq:rst-ic-v}) initial conditions, and the corresponding CMB angular $TT$ spectra. The details of how the latter kind of spectra are computed can be found in \cref{sec:results-cmb}. Note the dual $x$-axis shows both the wavenumber $k$ and the multipole $\ell$. The conversion between the two is performed using the Limber approximation, 
\begin{equation}\label{eq:limber-approx}
    \ell \approx k D_{\mathrm{rec}} = k \frac{r_s}{\theta_s},
\end{equation}
where $D_{\mathrm{rec}}$ is the distance to the last scattering surface at recombination, $r_s$ is the sound horizon at recombination, and $\theta_s$ is the angular parameter. The latter two can be derived from the Planck baseline CMB parameters listed in \cref{tab:cmbparams}. 

\begin{table}[htb!]
\renewcommand{\arraystretch}{1.5}
\begin{ruledtabular}
\caption{\label{tab:cmbparams} The fiducial values used for the cosmological parameters come from Planck~2018~TT,TE,EE+lowE~\cite{Planck2018}.}
\begin{tabular}{cccc}
parameter                    & fiducial value         & parameter         & fiducial value       \\\hline
$\Omega_\mathrm{b} h^2$      & $0.02236$              & $A_\mathrm{s}$    & $2.101 \times 10^{-9}$ \\
$\Omega_\mathrm{c} h^2$      & $0.1202$               & $n_\mathrm{s}$    & $0.9649$               \\
$h$                        & $0.6727$                & $\phi_\mathrm{i}$ & $\SI{16}{\planckmass}$ \\
$\tau_\mathrm{reio}$         & $0.0544$               & $N_\ast$          & $55$                   \\
\end{tabular}
\end{ruledtabular}
\end{table}

\begin{figure*}[tbp]
	\centering
	\subfloat[\label{fig:cut} Cutoff position.]{\includegraphics[scale=0.85]{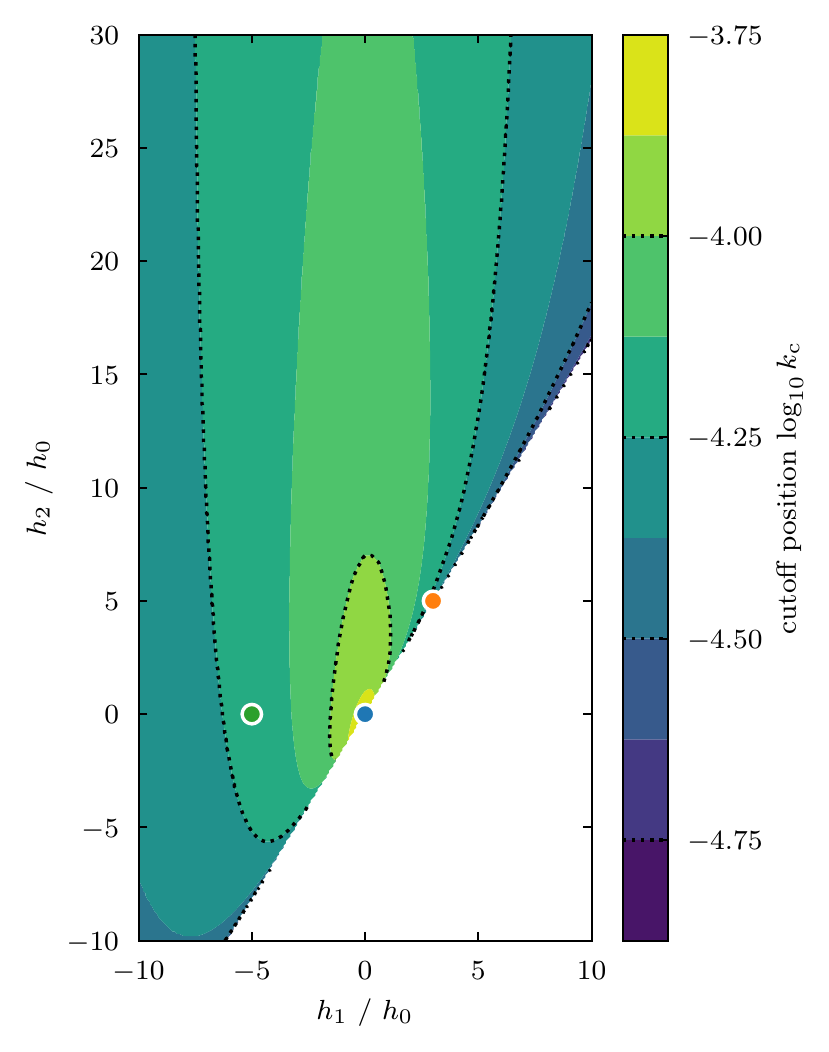}}
    \hfill
    \subfloat[\label{fig:amp} Amplitude of first oscillation.]{\includegraphics[scale=0.85]{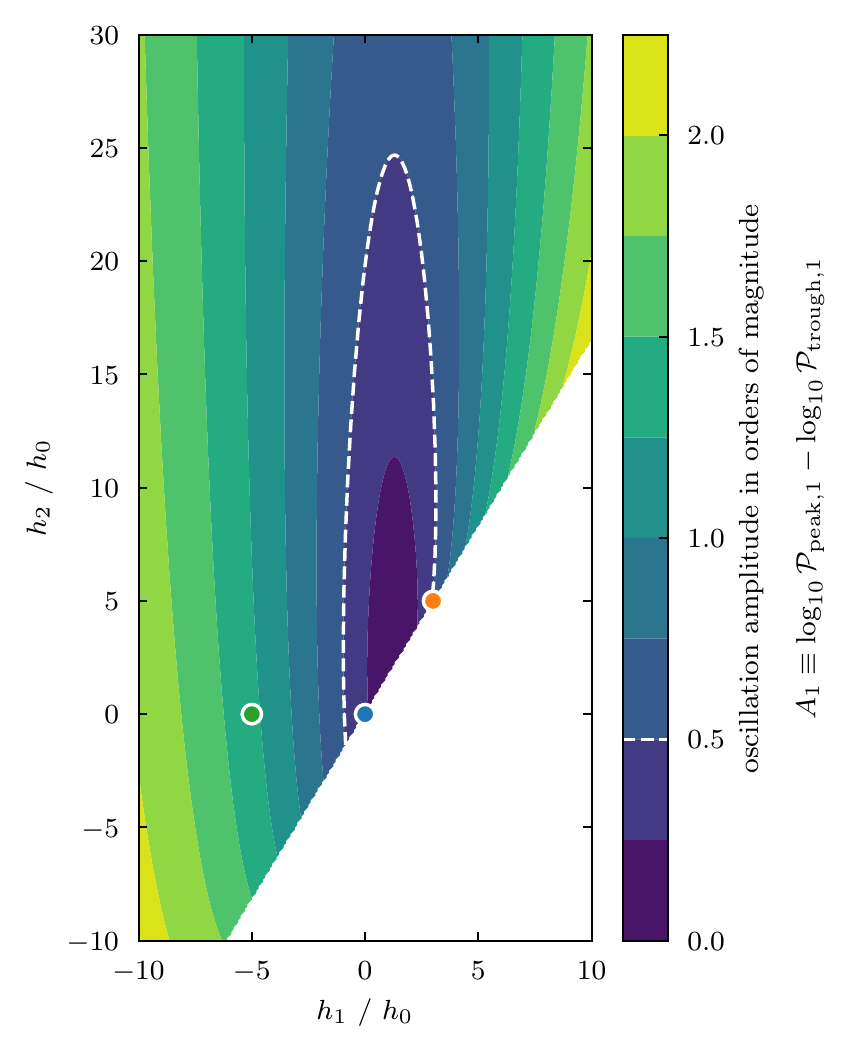}}
    \\
	\subfloat[\label{fig:avg} Average amplitude of first 10 oscillations.]{\includegraphics[scale=0.85]{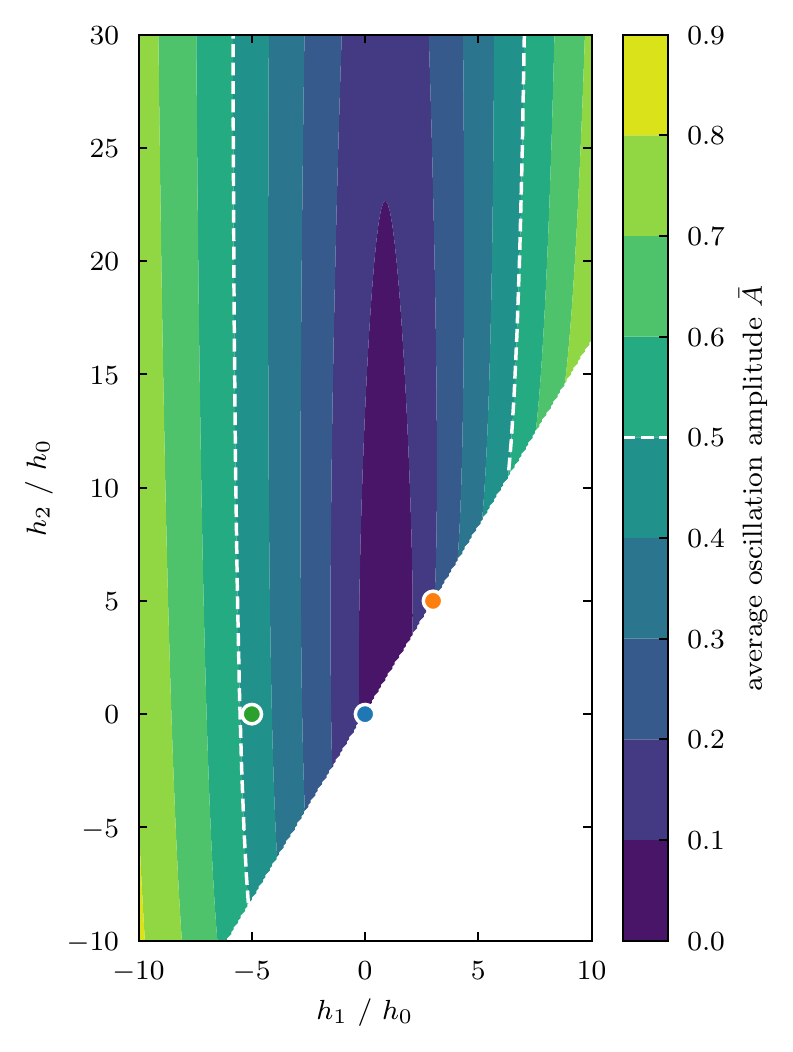}}
    \hfill
    \subfloat[\label{fig:rat} Ratio of first to second peak.]{\includegraphics[scale=0.85]{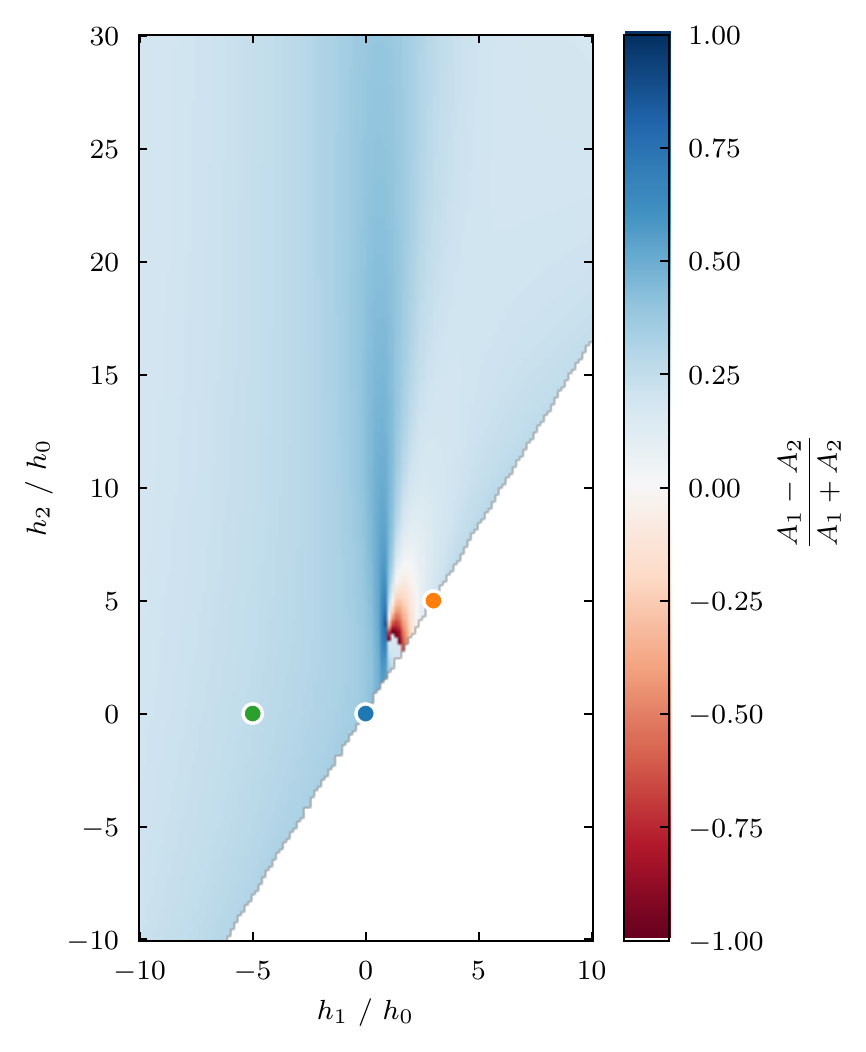}}
    \caption{\label{fig:contours}Dependence of features of the primordial power spectrum derived from Hamiltonian Diagonalisation initial conditions (\ref{eq:ic-hd-v-gen-dof1}) on the parameters $h_1/h_0$ and $h_2/h_0$. The cutoff position is measured as the wavenumber $k_\mathrm{c}$ where the power spectrum drops below a value of~$10^{-10}$ towards large scales. The oscillation amplitude is measured as the log-difference between a peak and the following trough. The three coloured points correspond to parameter pairs considered in \cref{sec:results-cmb}, their individual primordial power spectra are shown in \cref{fig:ppscmb}.}
\end{figure*}

\begin{figure*}[tbp]
	\hspace{1cm}
	\subfloat[][\label{fig:amp2} Amplitude of first oscillation. 
	
	(zoomed in from \cref{fig:amp})]{\includegraphics[scale=0.83]{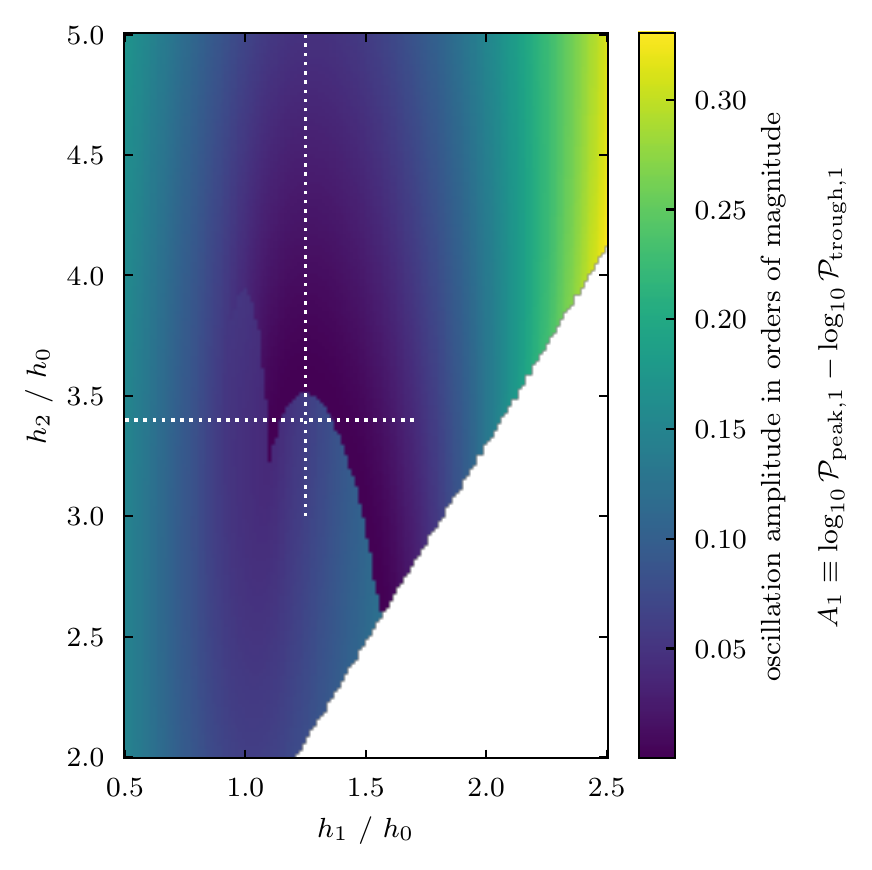}}
    \hfill
    \subfloat[][\label{fig:rat2} Ratio of first to second peak. 
    
    (zoomed in from \cref{fig:rat}) ]{\includegraphics[scale=0.83]{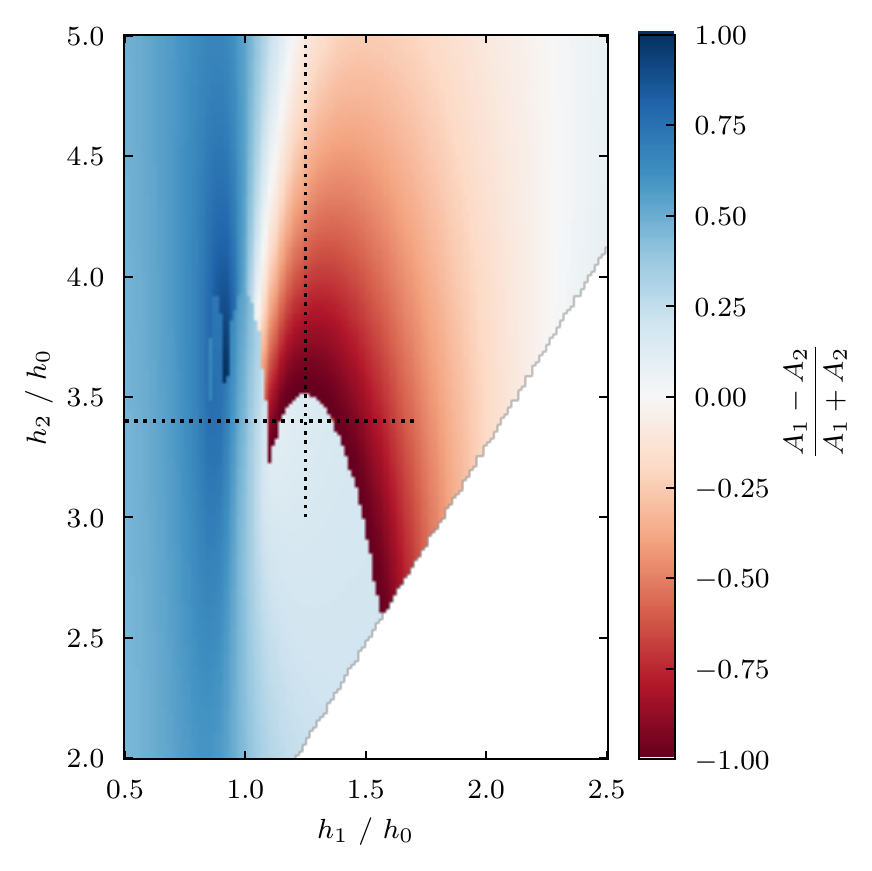}\hspace{1cm}}
    \\
	\subfloat[\label{fig:one}Sample spectra along horizontal line in upper panels.]{\includegraphics[width=0.49\textwidth]{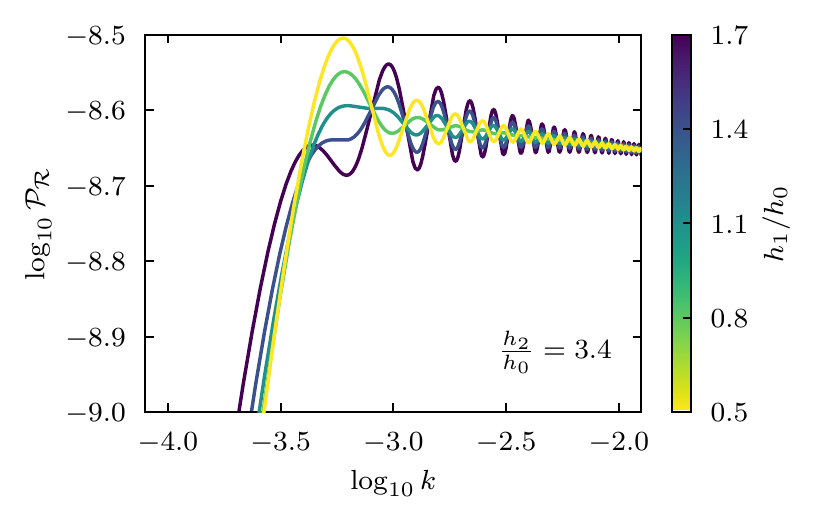}}
    \hfill
    \subfloat[\label{fig:two}Sample spectra along vertical line in upper panels.]{\includegraphics[width=0.49\textwidth]{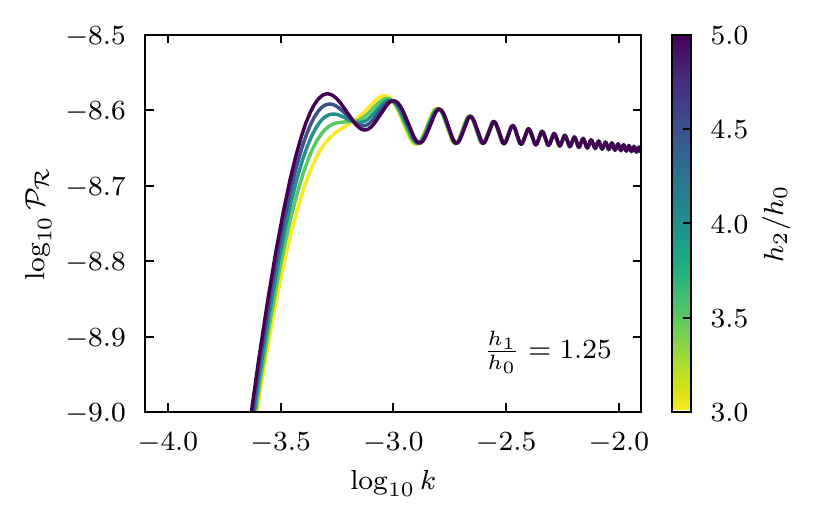}}
    \caption{\label{fig:contours2}Zoom-in of the centre region of the contour plots in \cref{fig:amp,fig:rat}. The dotted horizontal and vertical lines correspond to the line collections of primordial power spectra in the lower panels. The sharp lines in the contour plot correspond to the (dis-)appearance of the first peak.}
\end{figure*}

To show the full range of primordial power spectra achievable using Hamiltonian diagonalisation initial conditions (\ref{eq:ic-hd-v-gen-dof1}), we need to consider the two-dimensional space spanned by the two initial condition parameters, $h_2/h_0$ and $h_1/h_0$.
Since it is difficult to visualise individual spectra in this two-dimensional space, we opt to show instead contour plots of different features in the power spectra. 
The features considered are listed below.
\begin{enumerate}
    \item{\textbf{Cutoff position:} position of the low-$k$ cutoff present in all primordial power spectra considered. The cutoff is a result of considering kinetic dominance instead of eternal inflation, and is caused by perturbation modes that do not enter the Hubble horizon. We define the cutoff as the position where the amplitude of the power spectrum drops below~$10^{-10}$.}
    \item{\textbf{Amplitude of first oscillation:} calculated as the logarithmic difference between the first peak and the following trough.}
    \item{\textbf{Average amplitude of the first 10 oscillations.}}
    \item{\textbf{Ratio of the first to second peak:} defined as $ (A_1 - A_2)/((A_1 + A_2) $, where $A_{1,2}$ are the amplitudes of the first and second peaks, respectively.}
    \item{\textbf{Frequency of oscillations:} the leading frequency in the Fourier transform of $P_{\mathcal{R}}(k)$ was also considered as a feature initially. However, the frequency was not expected to change with $(h_1/h_0,h_2/h_0)$, which was indeed what has been found, and therefore the resulting contour plot is not shown.}
\end{enumerate}

\begin{figure}[tbp]
	\centering
    \includegraphics[]{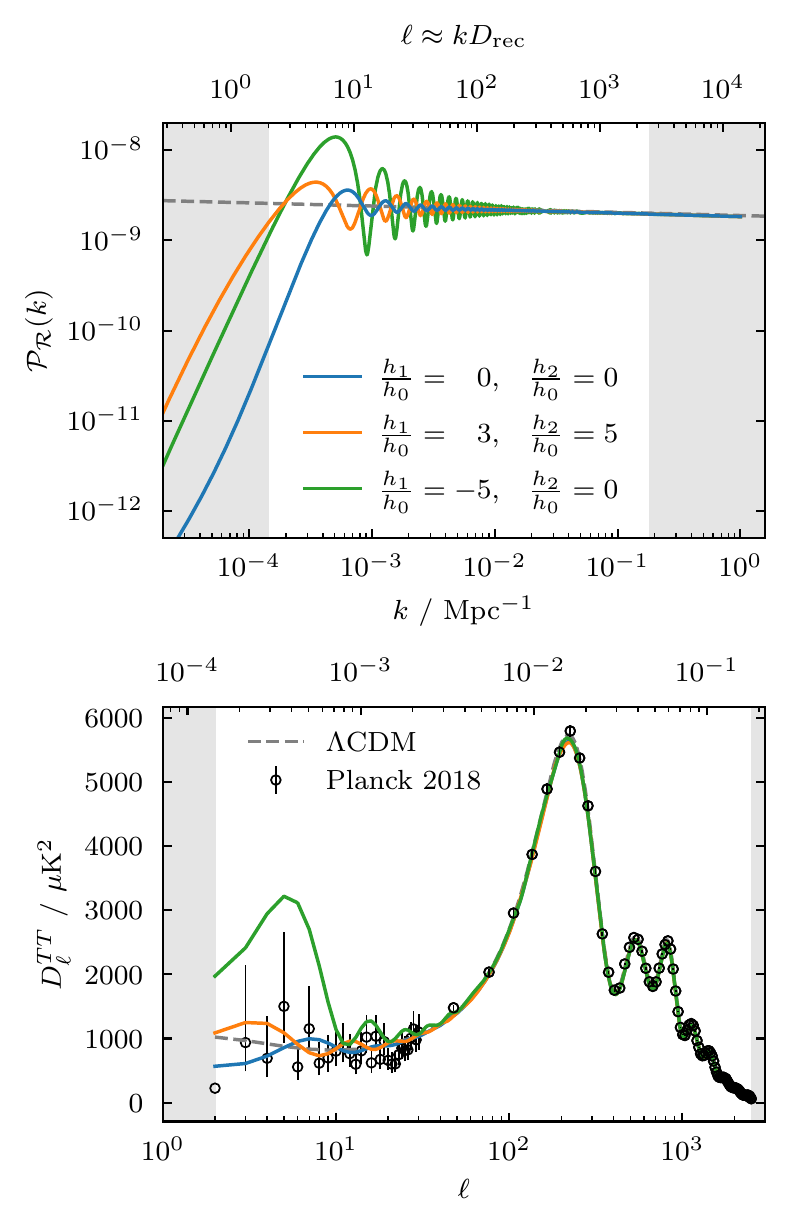}
    \caption{\label{fig:ppscmb}Primordial power spectrum (top panel) and angular CMB power spectrum of temperature anisotropies (bottom panel) for three different combinations of $h_1/h_0$ and $h_2/h_0$ as used in (\ref{eq:ic-hd-v-gen-dof1}).}
\end{figure}

\begin{figure}[tbp]
	\centering
    \includegraphics[]{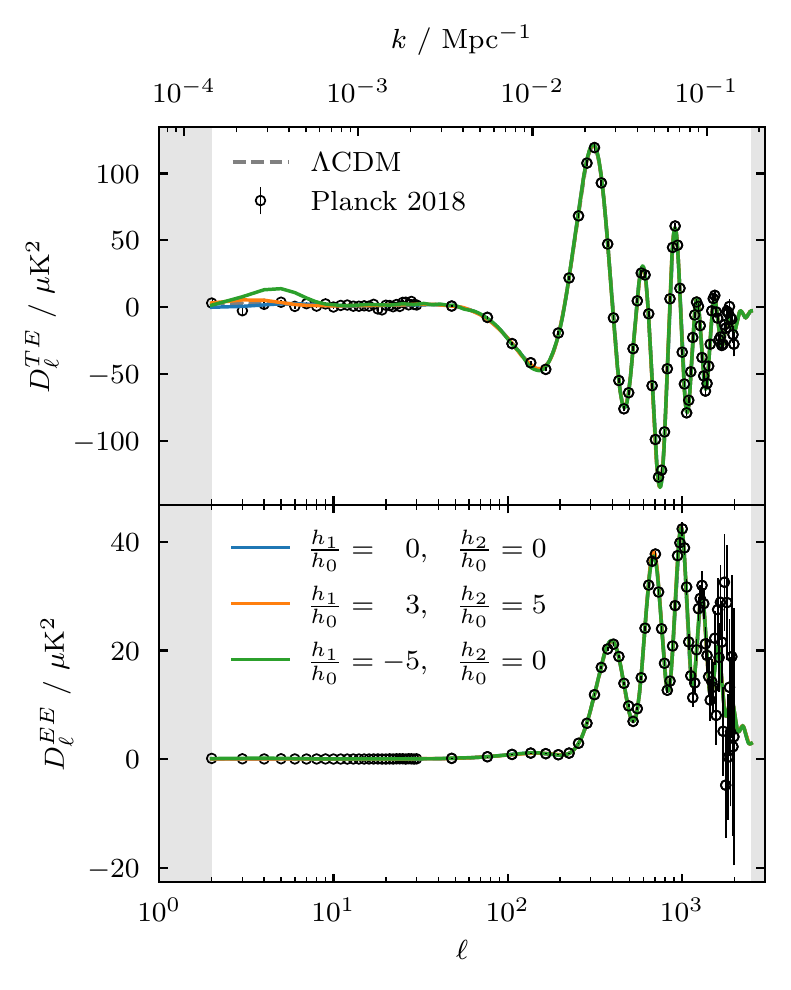}
    \caption{\label{fig:teee}CMB angular power spectra of the TE cross-spectrum and the EE auto-spectrum for three different combinations of $h_1/h_0$ and $h_2/h_0$ as in \cref{fig:ppscmb}.}
\end{figure}

\begin{figure}[tbp]
	\centering
    \includegraphics[]{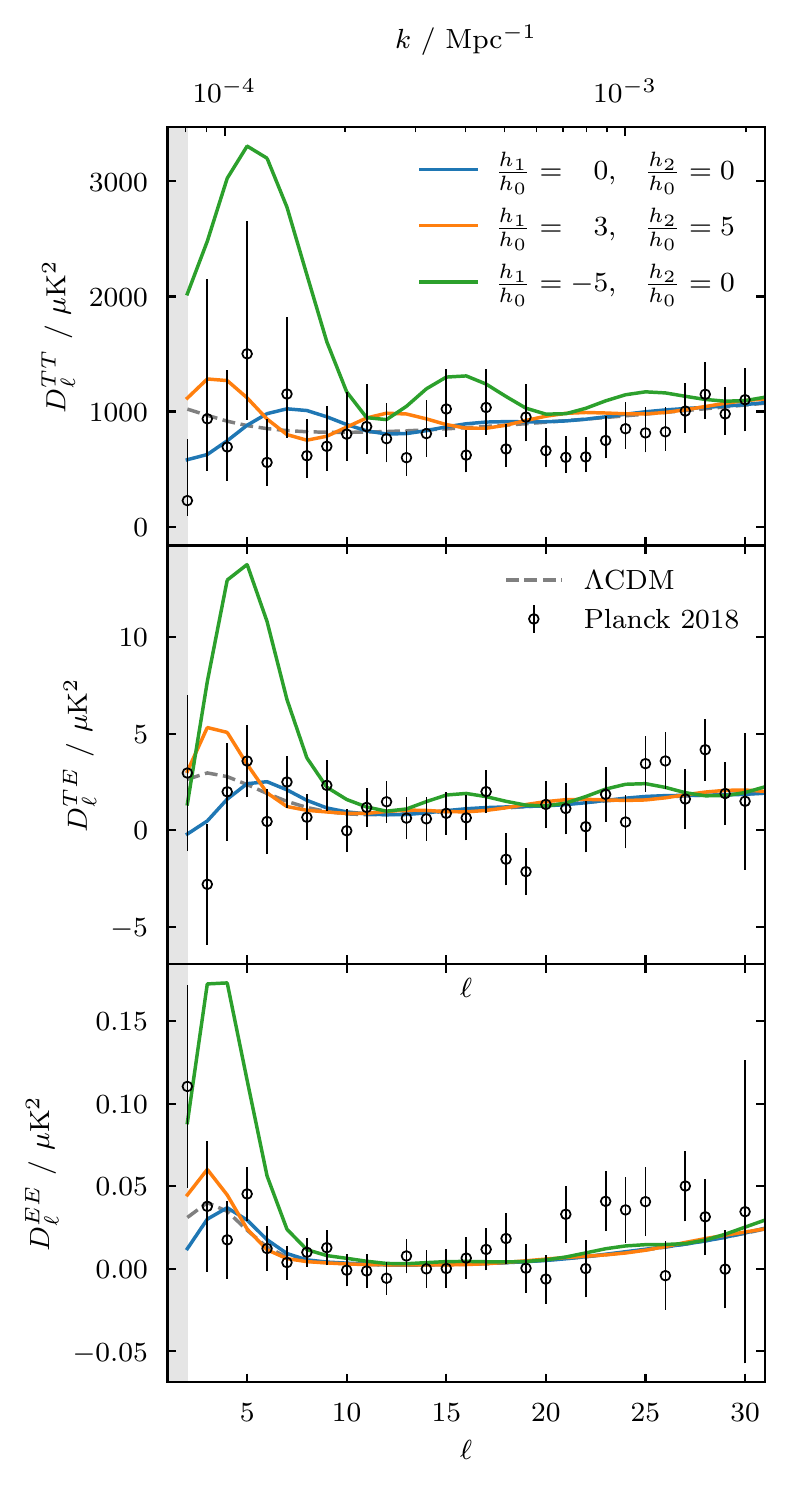}
    \caption{\label{fig:lowl}Low-$\ell$ region of the CMB angular power spectra from \cref{fig:ppscmb,fig:teee}.}
\end{figure}

\cref{fig:cut,fig:amp,fig:avg,fig:rat} show contour plots of the features 1--4 in primordial power spectra in $(h_1, h_2)$-space, with the three coloured points corresponding to parameter pairs considered in \cref{sec:results-cmb}. The empty `forbidden' region in all plots is a result of $h_2/h_0$ and $h_1/h_0$ appearing under a square root in (\ref{eq:ic-hd-v-gen-dof1}), causing there to be a lower limit to possible $k$ values if
\begin{equation}\label{eq:forbidden-region}
\frac{h_2}{h_0} + 2\frac{h_1z'}{h_0z} < 0.
\end{equation}
\cref{fig:amp,fig:rat} exhibit an interesting feature around $h_1/h_0 = 1.5 $, $h_2/h_0 = 3.5 $: the emergence of a new peak from the low-$k$ region of the primordial power spectrum. The phenomenon is shown in \cref{fig:amp2,fig:rat2,fig:one,fig:two}: \cref{fig:amp2} and \cref{fig:rat2} mark the region of interest in parameter space and the extent of inversion, while \cref{fig:one} and \cref{fig:two} show the emergence of the new peak along orthogonal directions in parameter space.

It is to be noted that one can arrive at the renormalised stress--energy tensor initial conditions (\ref{eq:rst-ic-v}) from Hamiltonian diagonalisation, by choosing the field to be quantised as $\mathcal{R}$, i.e.\ choosing $h=1$. This choice translates to $h_1/h_0=0,\:h_2/h_0=0$ in \cref{fig:contours,fig:contours2,fig:lowT_scatter}. The `standard' Danielsson prescription leads to the same initial conditions as minimising the renormalised stress--energy tensor, and so also corresponds to $h_1/h_0=0,\:h_2/h_0=0$. Conventional Hamiltonian diagonalisation is equivalent to choosing $h = z^{-1}$, or $h_1/h_0 \approx 0.82,\: h_2/h_0\approx1.37$ in \cref{fig:contours,fig:contours2,fig:lowT_scatter}.

\subsection{Cosmic Microwave Background}\label{sec:results-cmb}

\begin{figure}[tbp]
	\centering
    \includegraphics[scale=0.85]{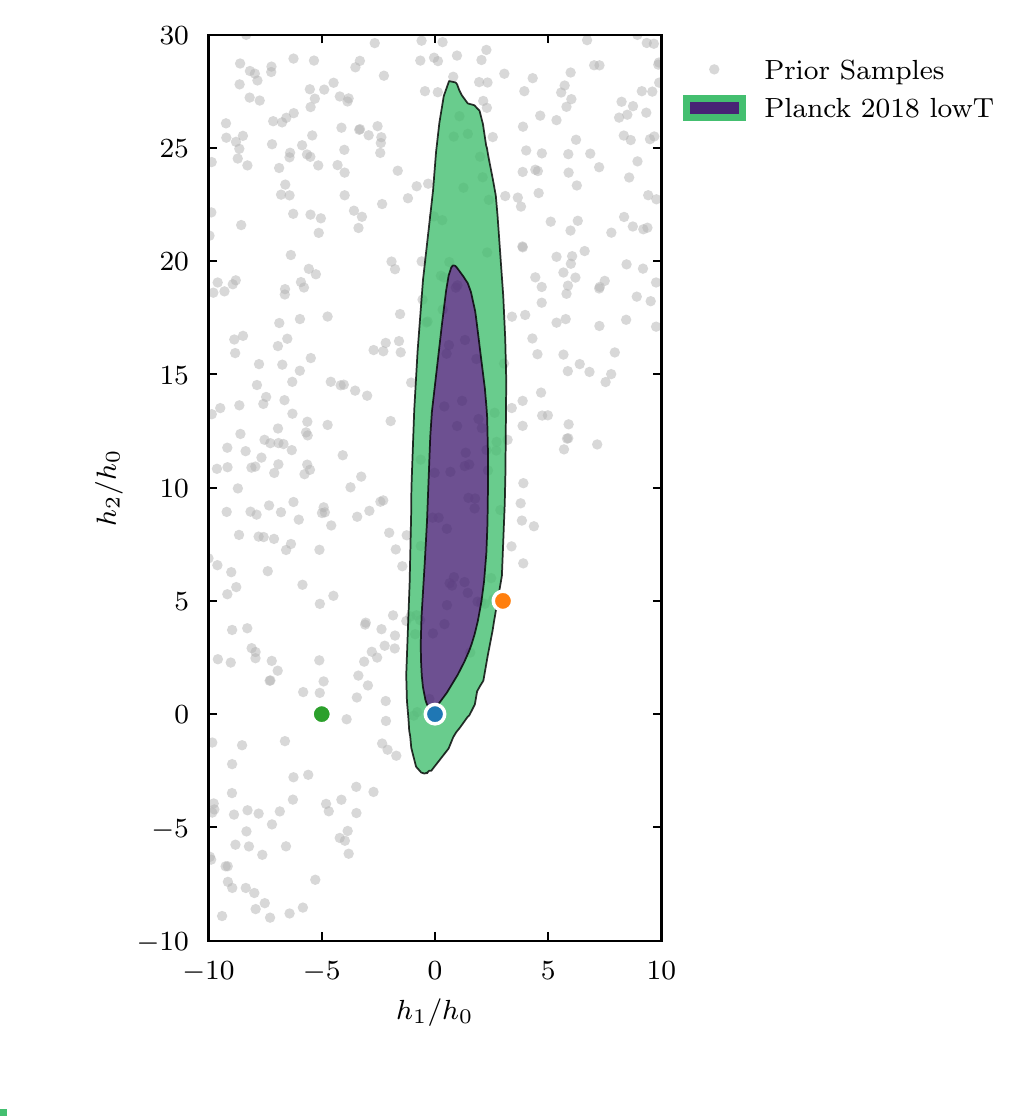}
    \caption{\label{fig:lowT_scatter} Prior samples and posterior contours showing the best-fit values of the $(h_1/h_0,h_2/h_0)$ parameters, given Planck 2018 lowT data. The coloured dots refer to parameter pairs described in \cref{sec:results-cmb} and \cref{fig:ppscmb}.}
\end{figure}

We generate the angular CMB $TT$, $TE$, and $EE$ power spectra starting from primordial power spectra using the Boltzmann code \texttt{CLASS} \cite{class-1,class-2,class-3,class-4}, with Planck 2018 baseline \cite{Planck2018} cosmological parameters as presented in \cref{tab:cmbparams}.

There are no obvious choices of physical features in the range of angular spectra observable today that would allow their visualisation in contour plots, therefore without any claim to completeness, we show the primordial power spectra and the corresponding $TT$, $TE$ and $EE$ spectra of three points in $(h_1/h_0, h_2/h_0)$-space marked in colour in \cref{fig:cut,fig:amp,fig:avg,fig:rat,fig:amp2,fig:rat2,fig:one,fig:two}. The primordial power spectra and the $TT$ spectra are shown in \cref{fig:ppscmb}, while \cref{fig:teee} shows the $TE$ and $EE$ spectra. The low-$\ell$ region of all these CMB spectra are shown separately in \cref{fig:lowl}.

Immediately it is clear that in the `just enough inflation' case, the choice of the function $h$ (through the values $h_1/h_0$ and $h_2/h_0$) can influence the CMB we observe today, despite $h$ stemming from the choice of canonical variables.  
Again we must stress that in the eternal inflation case, the different initial conditions discussed would yield the same observable spectra.

The question of whether observations could theoretically rule out certain regions of the $(h_1/h_0, h_2/h_0)$ parameter space is worth investigating, even though the inflationary potential used in this work, $V(\phi)\propto \phi^2$ has mostly been ruled out by the Planck 2015 results \cite{Planck2015infl}, and is only used here for the sake of simplicity. The posterior probability of $(h_1/h_0, h_2/h_0)$ given the Planck 2018 low-$\ell$ ($\ell \leq 30$) temperature-only likelihood~\cite{Planck2018likelihoods} is shown in \cref{fig:lowT_scatter}, where a uniform prior was used over the range $ -10 \leq h_1/h_0 \leq 10 $, $ -10 \leq h_2/h_0 \leq 30 $, excluding the forbidden region. The posterior has been explored using the PolyChord sampler~\cite{PolyChord1,PolyChord2}. The lowT data was used because \cref{fig:ppscmb,fig:teee} suggest that it is the low-$\ell$ region the different initial conditions have the most effect on.

It is clear that observational data can be used to select a preferred region of the initial condition parameters, but the point we would like to emphasise in this section is the wide-range effects that the choice of canonical variables can have on observations when using Hamiltonian diagonalisation to define the vacuum.

\section{Conclusions}\label{sec:conclusion}

In part, this work serves to highlight some points made by \citet{Fulling1979} regarding the lack of robustness of Hamiltonian diagonalisation under canonical transformations, in that we have shown that this procedure yields physically distinct vacua, and different initial conditions for the scalar curvature perturbations. We have shown that all sets of initial conditions derived this way can be cast into the general form proposed by \cite{Lim}, and therefore could be argued to be physically sensible. We have however illustrated the effect that the choice of initial conditions has on the primordial power spectrum and the CMB angular power spectra under the `just enough inflation' assumption. We showed that in such inflation models, different choices of canonical variables in a canonically non-invariant vacuum prescription such as Hamiltonian diagonalisation would be distinguishable by observation, noting that in a model where all perturbation modes spend a significant amount of time within the Hubble horizon, all initial conditions considered would yield the same observable results. We subjected two other choices of vacuum to the same transformations: one obtained by minimising the $00$-component of the renormalised stress--energy tensor \cite{nqicfi}, and the Danielsson vacuum \cite{danielsson}. In doing so we found that the Danielsson vacuum is inherently sensitive to canonical field-redefinitions and addition of surface terms, but the renormalised stress--energy tensor minimisation is not, despite not having been constructed to be invariant in this manner.

Many pieces of work since \citet{Fulling1979} acknowledge the existence of Hamiltonians related via canonical transformations, but suggest that there is a preferred Hamiltonian which leads to the conventionally used initial conditions. We noted that the initial conditions derived via the renormalised stress--energy tensor, which have thus been shown to be invariant under canonical transformations, differ from the initial conditions associated with this preferred Hamiltonian. Since the procedure of minimising the renormalised stress--energy tensor yields unambiguous initial conditions for the primordial perturbations and unambiguous quantum mechanical observables (expectation values of operators) under transformations that leave other physical traits (such as the equation of motion and commutator structure) invariant, we recommend it as the preferred prescription for setting the vacuum. 

\section*{Acknowledgements}

This work was performed using resources provided by the Cambridge Service for Data Driven Discovery (CSD3) operated by the University of Cambridge Research Computing Service (\url{www.csd3.cam.ac.uk}), provided by Dell EMC and Intel using Tier-2 funding from the Engineering and Physical Sciences Research Council (capital grant EP/P020259/1), and DiRAC funding from the Science and Technology Facilities Council (\url{www.dirac.ac.uk}). FJA was supported by STFC. LTH thanks the Isaac Newton Trust and STFC for their supoort. WJH thanks Gonville \& Caius College for their continuing support via a Research Fellowship.

\bibliography{min_RST_refs}
\appendix

\section{Hamiltonian diagonalisation under field-redefinitions and addition of surface terms}\label{sec:hd-gen}

We present the details of deriving initial conditions via Hamiltonian diagonalisation for the variable pair $(\tau, \chi)$, related to cosmic time and the gauge-invariant curvature perturbation $\mathcal{R}$ by  (\ref{eq:field-redef-hd}). For the derivations to follow in general, our computational strategy can be summarised as:
\begin{enumerate}
    \item Write down the action.
    \item Perform the transformation in question, expressing the action in terms of the field of interest.
    \item Find the equation of motion.
    \item Find the condition for the equation of motion to be of an undamped harmonic oscillator (for the purposes of quantising the field).
    \item Find the momentum conjugate to the field; quantise both as quantum oscillators.
    \item Write down the commutators of the system, and derive conditions for these to be canonical (thus ensuring that the transformation itself is canonical).
    \item Write down the definition of vacuum, i.e.\ the quantity to be minimised, and minimise it with respect to the Fourier modes of the quantised field, subject to the conditions derived.
\end{enumerate}

\subsection{Field redefinition}\label{sec:hd-fields-gen}

Changing variables in the perturbed action (\ref{eq:pert-action}) leads to
\be\label{eq:pert-action-gen}
\begin{split}
S &=  \int d^3x d\tau \frac{C}{2}\Bigg[ (\partial_{\tau}\chi)^2 +
2\chi(\partial_{\tau}\chi)\frac{\partial_{\tau}h}{h} \\
&+ \chi^2\left( \frac{\partial_{\tau}h}{h}\right)^2 - (\nabla \chi)^2\left(
\frac{h^2z^2}{C} \right)^2 \Bigg],
\end{split}
\ee
with $C = h^2\dot{\tau}az^2$.
This will in general yield the equation of motion of an oscillator with a non-zero first-derivative term. If the field $\chi$ is to be quantised as an oscillator, the first-derivative term needs to vanish, which holds if
\be\label{eq:c0-hd}
C = C_0 = \text{const}.
\ee
Setting $C$ to be constant, the equation of motion in terms of the Fourier modes of $\chi$ becomes
\be\label{eq:eom-chi}
0 = \partial_{\tau\tau}\chi_k + \Bigg[ \left( \frac{kh^2z^2}{C_0}\right)^2 + \frac{\partial_{\tau \tau}h}{h} - 2\left( \frac{\partial_{\tau}h}{ h} \right)^2 \Bigg]\chi_k.
\ee
To follow the conventional procedure, we integrate (\ref{eq:pert-action-gen}) by parts to eliminate the term linear in $\partial_{\tau}\chi$ to get
\be
\begin{split}
S =  \int d^3x d\tau \frac{C_0}{2}\Bigg[ &(\partial_{\tau}\chi)^2 + \chi^2\Bigg(\frac{\partial_{\tau\tau}h}{h}
 - 2\left( \frac{\partial_{\tau}h}{h}\right)^2 \Bigg) \\ - &(\nabla \chi)^2\left(
\frac{h^2z^2}{C_0} \right)^2 \Bigg].
\end{split}
\ee
The momentum conjugate to $\chi$ is then 
\be\label{eq:chi-conj-mom}
\pi_{\chi} = C_0 \partial_{\tau}\chi.
\ee 
We now quantise $\chi$ in the same way as $v$ in (\ref{eq:operator-v}). Imposing canonical commutation relations on the creation and annihilation operators results in the constraint
\be\label{eq:norm-chi}
(\partial_{\tau} \chi_k)\chi_k^{\ast} - (\partial_{\tau}\chi_k^{\ast})\chi_k = -\frac{i}{C_0},
\ee
because the momentum conjugate to $\chi$ was scaled by $C_0$. This is a necessary condition for the field redefinition to be a canonical transformation.
The normal-ordered Hamiltonian then takes the general form
\be\label{eq:hamiltonian-chi}
\begin{split}
H = \tfrac{1}{2}\int & \frac{\mathop{d^3k}}{(2\pi)^3}\Big[ \hat{a}_k\hat{a}_{-k}F_k(\tau) + \hat{a}_k^{\dagger}\hat{a}_{-k}^{\dagger}F_k^{\ast}(\tau) \\
&+ \big( 2\hat{a}_k^{\dagger}\hat{a}_k + \delta^{(3)}(0) \big)E_k(\tau) \Big],
\end{split}
\ee
with 
\be\label{eq:Ek}
\begin{gathered}
E_k(\tau) = |\partial_{\tau}\chi_k|^2 + \omega_k^2|\chi_k|^2,\\
F_k(\tau) = (\partial_{\tau}\chi_k)^2 + \omega_k^2\chi_k^2, \\
\omega_k^2 = \left( \frac{kh^2z^2}{C_0}\right)^2 + \frac{\partial_{\tau \tau}h}{h} - 2\left( \frac{\partial_{\tau}h}{
h} \right)^2.
\end{gathered}
\ee
Minimising the vacuum expectation value of the Hamiltonian for each $k$-mode separately, subject to the constraint (\ref{eq:norm-chi}) thus gives the generalised initial conditions
\be\label{eq:ic-hd-chi-appendix}
\begin{split}
|\chi_k|^2 &= \left(2C_0\omega_k\right)^{-1},\\
\partial_{\tau}\chi_k &= -i\omega_k\chi_k.
\end{split}
\ee

\subsection{Surface terms}\label{sec:hd-surfterms-gen}

In order to get to the form of the Hamiltonian (\ref{eq:hamiltonian-chi}), we integrated by parts once and discarded a boundary term:
\be\label{eq:b-term-chi} 
\begin{split}
\mathcal{S} &\supset \int \mathop{d^3x}\mathop{d\tau} \frac{C_0}{2}\Bigg[ 2\chi (\partial_{\tau}\chi) \frac{\partial_{\tau}h}{h} + \chi^2\left(\frac{\partial_{\tau}h}{h}\right)^2  \Bigg] \\
&= \Bigg[ \chi^2\frac{\partial_{\tau}h}{h}\Bigg]
- \int \mathop{d^3x}\mathop{d\tau}\Bigg[ \frac{\partial_{\tau \tau}h}{h}\chi^2- 2\left( \frac{\partial_{\tau}h}{h}\right)^2\chi^2 \Bigg].
\end{split}
\ee
This is equivalent to adding the surface term
\be
-\partial_{\tau}\left( \chi^2 \frac{\partial_{\tau} h}{h} \right).
\ee
If we instead kept the action in its original form, (\ref{eq:pert-action-gen}), the momentum conjugate to $\chi$ would be
\be
\pi = C_0\left( \partial_{\tau}\chi + \frac{\partial_{\tau}h}{h}\chi \right),
\ee
and the normal-ordered Hamiltonian would take the same form as (\ref{eq:hamiltonian-chi}), but now with
\be\label{eq:hamiltonian-chi-bt}
\begin{gathered}
E_k(\tau) = |\partial_{\tau}\chi_k|^2 + \sigma_k^2|\chi_k|^2, \\
F_k(\tau) = (\partial_{\tau}\chi_k)^2 + \sigma_k^2\chi_k^2, \\
\sigma_k^2 = \left( \frac{kh^2z^2}{C_0} \right)^2 - \left(\frac{\partial_{\tau}h}{h}\right)^2.
\end{gathered}
\ee
As seen by symmetry, this yields initial conditions for $(\tau, \chi)$ of the same form as (\ref{eq:ic-hd-v-gen}), but with $\omega_k$ replaced with $\sigma_k$. Switching back to $(\eta, v)$, they are
\be\label{eq:hd-ic-v-gen-bt-appendix}
\begin{gathered}
|v_k|^2 = \frac{1}{2\sqrt{k^2 - \left(\frac{h'}{h}\right)^2}},\\
v_k' = \Bigg( -i\sqrt{ k^2 - \left(\frac{h'}{h}\right)^2 } + \frac{h'}{h} + \frac{z'}{z} \Bigg)v_k.
\end{gathered}
\ee

\section{Minimising the renormalised stress--energy tensor under field redefinitions}\label{sec:rst-gen}

We generalise the results (\ref{eq:rst-ic-y})--(\ref{eq:rst-ic-v}) in analogy with \cref{sec:hd-fields-gen}. We rewrite the action (\ref{eq:action-scalar}) in terms of the redefined time and field
\be\label{eq:field-redef-rst-appendix}
t \rightarrow \tau, \quad \varphi \rightarrow \chi(x, \tau) \equiv \frac{\varphi}{h} = \frac{y}{ah},
\ee
keeping in mind that the metric then changes to $g_{\mu \nu} = \text{diag}(\dot{\tau}^{-2}, -a^2, -a^2, -a^2)$. This yields
\be\label{eq:action-scalar-gen}
\begin{split}
S = \int & \mathop{d^3x}  \mathop{d\tau} \frac{C}{2}\Bigg[ (\partial_{\tau}\chi)^2 + 2(\partial_{\tau}\chi)\chi\frac{\partial_{\tau}h}{h} \\
&+ \chi^2 \Big( \frac{\partial_{\tau}h}{h}\Big)^2 - (\nabla \chi)^2\Big(\frac{h^2a^2}{C_0}\Big)^2 \\
&- \chi^2m^2\Big( \frac{h^2a^3}{C_0} \Big)^2 \Bigg],
\end{split}
\ee
with $C = h^2\dot{\tau}a^3$. As before, the equation of motion resulting from the variation of this action can be made first-derivative-free by setting
\be
C = C_0 = \text{const},
\ee
in which case it becomes, in Fourier space,
\be\label{eq:eom-chi-scalar}
\begin{split}
& \partial_{\tau\tau}\chi_k + \Bigg[ \left(\frac{kh^2a^2}{C_0}\right)^2 + \frac{\partial_{\tau\tau}h}{h} \\
&- 2\left(\frac{\partial_{\tau}h}{h}\right)^2 + \left(\frac{mh^2a^3}{C_0}\right)^2 \Bigg]\chi_k = 0.
\end{split}
\ee

The field $\chi$ is then quantised, which makes $\varphi$ take the form
\be\label{eq:operator-chi-rst}
\varphi (x) = \int \frac{d^3k}{(2\pi)^3}h(\tau) \Big[ \hat{a}_k
\chi_k(\tau) e^{i\vec{k}\cdot\vec{x}} +\hat{a}_k^{\dagger}\chi_k^{\ast}(\tau)e^{-i\vec{k}\cdot\vec{x}}
\Big].
\ee
The Hadamard Green function can then be written
\be
\begin{split}
G^{(1)}(x, x') &= \int \frac{d^3k}{(2\pi)^3} h(\tau)h(\tau') \Big[ \chi_k(\tau)\chi_k^{\ast}(\tau')e^{i\vec{k}\cdot(\vec{x}-\vec{x'})} \\
&+ \chi_k^{\ast}(\tau)\chi_k(\tau')
e^{-i\vec{k}\cdot(\vec{x}-\vec{x'})} \Big].
\end{split}
\ee
Acting on this with the bi-scalar derivative function and taking the $00$ component thus gives
\be\label{eq:rst-appendix}
\begin{split}
\bra{0}T_{00}\ket{0}_{\text{ren}} &= \tilde{T} + \frac{1}{2} \int \frac{d^3k}{(2\pi)^3}
h^2\Big[\Big( \partial_{\tau}\chi_k + \frac{\partial_{\tau}h}{h}\chi_k \Big) \\
&\cdot \Big( \chi_k^{\ast}{} +\frac{\partial_{\tau}h}{h}\chi_k^{\ast}\Big) +\Big( \frac{k^2}{a^2\dot{\tau}^2} +
\frac{m^2}{\dot{\tau}^2}\Big)\chi_k \chi_k^{\ast}  \Big] .
\end{split}
\ee
The normalisation constraint on the mode functions $\chi_k$ takes the same form as for the generalised perturbed classical action in \cref{sec:hd-fields-gen}, and is therefore given by (\ref{eq:norm-chi}).
Minimising the expression (\ref{eq:rst-appendix}) subject to this constraint gives for the mode functions $\chi_k$ 
\be\label{eq:rst-ic-chi-appendix}
\begin{gathered}
|\chi_k|^2=\frac{1}{2\sqrt{h^4a^4\left(k^2 + m^2a^2\right)}},\\
\partial_{\tau}\chi_k = \Bigg( -\frac{i}{C_0}\sqrt{h^4a^4\left(k^2 + m^2a^2\right)}
-\frac{\partial_{\tau}h}{h}\Bigg)\chi_k.
\end{gathered}
\ee

\end{document}